\documentclass[twocolumn]{article}
\usepackage[english]{babel}
\usepackage{natbib}
\usepackage{amsmath}
\usepackage{graphicx}
\usepackage[hidelinks]{hyperref}{}
\usepackage{float}
\usepackage{txfonts}
\usepackage{amssymb}
\usepackage{mathrsfs}
\usepackage{stmaryrd}
\usepackage{multirow}
\usepackage{multicol}
\usepackage{subcaption}
\usepackage[labelfont=bf]{caption}
\usepackage{lmodern}
\usepackage{wrapfig}
\usepackage{blindtext}
\usepackage{enumitem}
\usepackage{amssymb}
\usepackage{esvect}
\usepackage{empheq}
\usepackage{cancel}
\usepackage{hhline}
\usepackage{natbib}
\usepackage{times}
\usepackage{color}
\usepackage{authblk}
\usepackage[margin=0.65in]{geometry}
\hypersetup{
    colorlinks=true,
    citecolor=blue,
    linkcolor=red,
    urlcolor=black
    }

\begin{document}

\title{A complete study of the precision of the concentric MacLaurin spheroid method to calculate Jupiter's gravitational moments}
% \titlerunning{Study of the CMS method}

\author[1,2]{F. Debras}
\author[1,2]{G. Chabrier}

\affil[1]{Ecole normale sup\'erieure de Lyon, CRAL, UMR CNRS 5574, 69364 Lyon Cedex 07,  France}
\affil[2]{School of Physics, University of Exeter, Exeter, EX4 4QL, UK}

\date{\today}
\twocolumn[
\begin{@twocolumnfalse}
\maketitle Accepted in A\&A
\begin{abstract}
% \abstract
A few years ago, Hubbard (2012, 2013) presented
an elegant, non-perturbative method, called concentric MacLaurin spheroid (CMS), to calculate with very high accuracy the gravitational moments
of a rotating fluid body following a barotropic pressure-density relationship.
Having such an accurate method is of great importance for taking full advantage of the Juno mission, and its extremely
precise determination of Jupiter gravitational moments, to better constrain the internal structure of the planet. 
Recently,
several authors have applied this method to the Juno mission with 512 spheroids
linearly spaced in altitude. We demonstrate in this
paper that such calculations lead to errors larger than Juno's error bars, invalidating 
the aforederived Jupiter models at the level required by Juno's precision. 
We show that, in order to fulfill Juno's observational constraints, at least 1500 spheroids must be 
used with a cubic, square or exponential repartition, the most reliable solutions. 
When using a realistic equation of state 
instead of a polytrope, 
we highlight the necessity to properly describe 
the outermost layers to derive an accurate boundary condition, excluding in particular a zero pressure outer condition. 
Providing all these constraints are fulfilled,
the CMS method can indeed be used to derive Jupiter models within Juno's present
observational constraints. However, we show that the treatment of the outermost layers leads to 
irreducible errors in the calculation of the gravitational moments and thus on the inferred physical quantities 
for the planet. We have quantified these errors and evaluated the 
maximum precision that can be reached with the CMS method in the present and future exploitation of Juno's data.

\end{abstract}
\vspace{0.5cm}
\end{@twocolumnfalse}
]

% \keywords{Planets and satellites: gaseous planets -- Planets and satellites: interiors -- 
% Equation of state -- Methods: numerical}

\section{Introduction} \label{sec:intro}

The concentric Maclaurin spheroid (CMS) method has been developed 
by Hubbard in
 \cite{Hub2012} and \cite{Hub2013} (H13). 
This method consists of a numerical hydrostatic scheme which decomposes a rotating celestial body into
$N$ spheroids of constant density (the well-known MacLaurin spheroid would correspond to the case $N=1$). It needs two inputs. First, (i)
the planet rotational distortion, $q=\omega^2a^3/GM$, where $\omega$ denotes the angular velocity, $a$ the equatorial radius, $M$ the planet's mass and $G$
is the gravitational constant. The factor $q$, which is linked 
to the MacLaurin's parameter, $m=3\omega^2/(4\pi\rho G)$, is also used in the theory of Figures
(see for example \cite{Zharkov} or \cite{Chab92}),
and represents the ratio of the rotational over gravitational potentials. Second, (ii) a 
barotrope ($P$,$\rho$) representing the equation of state (eos) of the planet and its composition. 

For a given density profile,
we can calculate the gravitational potential of each spheroid (which is constant on the 
spheroid), and then calculate self-consistently the radius of each spheroid as a function of latitude. 
The obtained radii give different values for the potential of the spheroids, 
and successive iterations lead to convergence of both shape and potential of the 
spheroids (see \cite{Hub2012} and H13). With the density and potential of each spheroid,
one easily obtains the pressure from the hydrostatic equilibrium condition. 

\noindent After this first stage of convergence, it is necessary to verify whether the obtained ($P$,$\rho$) profile 
of the planet is in agreement with the prescribed barotrope. Since generally it is not, 
an outer loop is necessary to converge the density profile for the given
eos. 

At last, once the radii, potentials and densities of the multi-layer spheroids
have been obtained consistently with the required numerical precision, one obtains
a discretized profile of these quantities throughout the whole planet.
We can then calculate the gravitational moments, to be compared with Juno's
observations. The formula is, by additivity: 
\begin{equation}
J_k^{ext} = \sum_{i=0}^{N-1} J_k^i \times  \left(\lambda_i\right)^k,
\label{J}
\end{equation}
where $J_k^{ext}$ is the moment of order $k$, $J_k^{i}$ is the 
moment of order k of the $i^{th}$ spheroid, $\lambda_i$ is the ratio of
the equatorial radius of the $i^{th}$ spheroid over the equatorial radius 
of the planet and the spheroids have been labeled with index $i=0,N-1$, with $i=0$ corresponding to 
the outermost spheroid and $N-1$ to the innermost one. Eq.(\ref{J}) 
implies that the uncertainties of each spheroid are added up, 
so a small error on the radius, potential or density can lead ultimately to a significant error
once summed over all the layers. 

As explained in H13, for a given piecewise density profile, the evaluation
of the gravitational moments is extremely precise, with an error around 
$10^{-13}$. The main sources of error, apart from the uncertainties on the barotrope,
then arise from the 
finite number of spheroids and from the approximation of constant density within each spheroid, potentially leading 
to an incorrect evaluation of the radii and potential of the layers. In this paper, we
evaluate the errors due to the discretization of the density distribution, and of the description
of the outermost layers, the main aim of the paper is to find the best configuration in the CMS method to safely use it in the context
of the Juno mission.

In Section \ref{analytics}, we show analytically that 
512 spheroids are not enough to match Juno's error bars, 
and that the repartition of spheroids is an important parameter in the evaluation of the errors.
We confirm these results by numerical
calculations in Section \ref{numerics}. We then apply our method with
a realistic eos in Section \ref{sec:SCVH}. These sections
show the need to improve the basic CMS method.
In Section \ref{sec:BC}, we study the impact of the outer boundary condition.
Section \ref{conclusion} is devoted to the conclusion.

\section{Analytical evaluation of the errors in the CMS method}
\label{analytics}
\subsection{Evaluation of the errors with 512 spheroids}

% \startlongTable
\begin{table*}
\captionsetup{labelfont=bf}
\caption{Values of the planetary parameters of Jupiter. $R_{eq}$ and $R_{polar}$
are observed at 1 bar. The value of the pulsation is chosen following
\cite{Archinal2011}.
(a) \cite{Cohen87}
(b) \cite{Folkner2017} (c) \cite{Archinal2011}
(d) \cite{Riddle76}
(e) \cite{bolton2017}}
\label{values_Jupiter}
\centering
\begin{tabular}{ll}
\hline
\hline
Parameter & \multicolumn{1}{c}{Value} \\
\hline
$G$$^a$ (global parameter) & $6.672598 \cdot 10^{-11} \pm 2 \cdot 10^{-17}$ m$^3$kg$^{-1}$s$^{-2}$ \\
$G*M_J$$^{b}$ &  $126 686 533 \pm 2 $ $\cdot 10^9$m$^3$s$^{-2}$\\
$M_J$ & $1.89861 \cdot 10^{27}$ kg \\
$R_{eq}$$^c$ & $71492 \pm 4$ km \\
$R_{polar}$$^c$ & $66854 \pm 10$ km \\
$\omega$$^d$ & $1.7585324 \cdot 10^{-4}$ $\pm 6 \cdot 10^{-10}$ s$^{-1}$ \\
$\bar{\rho}$ & $1326.5 $ kg.m$^{-3}$ \\
$m = 3 \omega^2/4 \pi G \bar{\rho}$ & 0.083408 \\
$q = \omega^2 R_{eq}^3/G M_J$ & 0.0891954 \\
$J_2 \times 10^6$$^e$ & $14696.514 \pm 0.272$ \\
$-J_4 \times 10^6$$^e$ & $586.623 \pm 0.363$\\
$J_6 \times 10^6$$^e$ & $34.244 \pm 0.236$\\
$-J_8 \times 10^6$$^e$ & $2.502 \pm 0.311$\\
\hline
\end{tabular}
\end{table*}

In order to evaluate the errors analytically, we have considered a polytrope of index
$n=1$, $P\propto \rho^2$. We call $\rho_2$ the discretized version of $\rho$ 
in the CMS model. The various quantities relevant for Jupiter are given in Table \ref{values_Jupiter}.

If we assume longitudinal and 
north-south symmetry, only the even values of the gravitational moments are not null. 
Therefore, we were able to express the difference
between the analytical value of the potential and the value obtained
with the CMS method on a point exterior to the planet with radial 
and latitudinal coordinates $(r,\mu=cos(\theta)$, where $\theta$ is the angle from the rotation axis), as: 
\begin{gather}
\Delta \phi = -\dfrac{4\pi G}{r} \sum_{k=0}^{\infty}P_{2k}(\mu)
r^{-2k} \nonumber \\
\times \int_{0}^{1}\int_{0}^{a_J}\left[\rho (r') - \rho_2(r')  \right]
r'^{2k+2} P_{2k}(\mu) dr' d\mu, 
\label{potential_tot}
\end{gather}
where $P_{2k}$ is the Legendre polynomial
of order $2k$ and $a_J$ is Jupiter's equatorial radius. Our first assumption
here was that the CMS method prefectly captures the shapes of 
the spheroids. In reality, this is the case only for an infinite number of 
spheroids, but we have neglected this first source of error which is difficult
to evaluate analytically. 

\noindent Since the masses of the two models must be the same,
the $k=0$ term must cancel. 
In terms of gravitational moments
$J_{2k}$ :

\begin{equation}
J_{2k} = - \dfrac{4\pi}{Ma_J^{2k}}\int_{0}^{1}\int_{0}^{a_J}\rho (r')
r'^{2k+2} P_{2k}(\mu) dr' d\mu.
\label{def_J2k}
\end{equation}

\noindent By construction we can write
$\rho_2(r) = \rho_i ,\text{   } r_{i+1}<r\leq r_i$. So the difference on the gravitational moment between the exact 
and the CMS models from Eq.(\ref{def_J2k}) is:

\begin{gather}
\Delta J_{2k} = - \dfrac{4\pi}{Ma_J^{2k}}
\sum_{i=0}^{N-1}
\int_{0}^{1}\int_{r_{i+1}(\mu)}^{r_i(\mu)}
\left[\rho (r') - \rho_i   \right]  \nonumber \\
\,\,\,\,\,\,\,\,\,\,\times r'^{2k+2} P_{2k}(\mu) dr' d\mu. \label{eq:def_deltaJ2k}
\end{gather}

\noindent This formulation is exactly the same as in H13 (Eq.(10)),
except that we used $\rho_i$ instead of $\delta \rho_i$. The analytical expression
of the density of a rotating polytrope at first order in $m$ as a function of 
the mean radius $l$ of the equipotential layers and of the planet mean density 
$\bar{\rho}$ is given by (\cite{Zharkov}): 

\begin{gather}
\dfrac{\rho(l)}{\bar{\rho}} = A \dfrac{sin(\alpha \dfrac{l}{l_0})}{(\dfrac{l}{l_0})}
+ \dfrac{2}{3} m \label{density_ZT},\\
{\rm with}\,\, A = \dfrac{\pi}{3} \left( 1-\dfrac{2}{3}m \left( 1-\dfrac{6}{\pi^2} \right) \right)
\text{    and    } \alpha = \pi \left(1 + \dfrac{2m}{\pi^2} \right),
\end{gather}
where $l_0$ is the outer mean radius. \\

At this stage, our method implied an important approximation, namely that
$\rho_2$ followed perfectly the polytropic density profile of Eq.(\ref{density_ZT})
on each layer, if evaluated at the middle of the layer: 
$\rho_i = \rho_{poly} \left(r_i+r_{i+1}\right)/2 $.
The errors we calculated are thus smaller than the real ones since we did not take into account those arising from
the departure from the exact density profile.
In Appendix \ref{mu_legendre},
we show how to get rid of the dependence on the angular part, $\mu$, in Eq.(\ref{eq:def_deltaJ2k}). As detailed in 
Appendix \ref{maths}, for a linear spacing of the spheroid with depth, $\Delta r = a_J/N$, we obtained:

\begin{gather}
%\boxed{
|\Delta J_{2k}| \sim (2k+3)\dfrac{\pi}{12}\dfrac{<P_{2k}>}{N^3}
\sum_{i=0}^{N-1} \left(1 - \dfrac{1}{N}\left(i+\dfrac{1}{2} \right)\right)^{2k+2},
%} 
\label{J_2k_final} \\
{\rm with} <P_{2k}> = \bigg |\int_{0}^{1}
\left(\dfrac{1}{1+e^2\mu^2}\right)^{k+1} P_{2k}
(\mu)  d\mu \bigg |. 
\label{def_P2k}
\end{gather}

\begin{figure}[ht]
\includegraphics[width=1\linewidth]{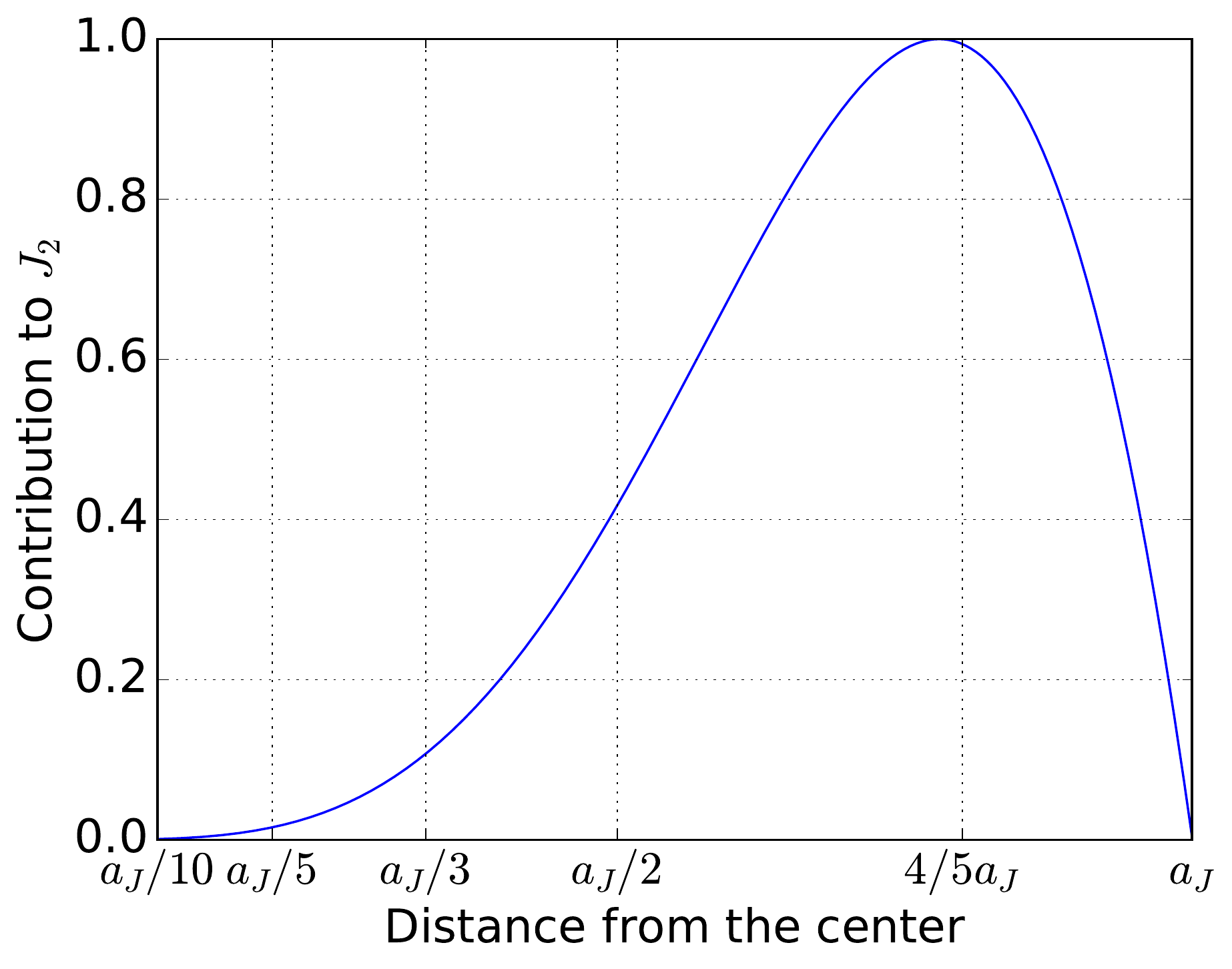}
\caption{Contribution of each part of the planet to the total
value of $J_2$, normalized to its maximum value, for a polytrope of index $n=1$.}
\label{impact_J2}
\end{figure}

It is interesting to note that the relative error of each layer is
a decreasing function of $i$: the external layers lead to larger
errors on the gravitational moment than the inner ones.
Figure \ref{impact_J2} displays 
the contribution to $J_2$ of the outer layers of the planet
in the polytropic case (we have
numerically integrated the expression for $J_2$ from 
Eq.(\ref{def_J2k}), Eq.(\ref{density_ZT}) and Eq.(\ref{def_<P>})). 

To quantify the error, we focused on $J_2$,  the case $k=1$.  
As detailed in Appendix \ref{maths_J2},
one gets:

\begin{equation}
%\boxed{
|\Delta J_{2}| \times 10^{6} \sim \dfrac{\pi}{12} \dfrac{<P_2>}{N^2}\times 10^{6} 
 \approx \dfrac{9.4\cdot 10^{3}}{N^2}.
%}
\label{J_2_fin}
\end{equation}
With $N=512$ (as used in H13), we get $|\Delta J_{2}| \times 10^{6} \sim 3.6 \times 10^{-2}$. The first data
from Juno after three orbits give an uncertainty of $2.72 \times 10^{-1}$ on $J_2\times 10^{6}$.
We are under these error bars by a factor of eight, but in these analytical calculations, we have made
several restrictive approximations. Notably, we have supposed that the numerical
method can find perfectly the shape and the density of each spheroid, 
and we have neglected the factor $\alpha^3/3\sim 2-10$ in Eq.(\ref{C1_C2}). 
As mentioned earlier, it is hard to evaluate a priori the errors due to the wrong
shapes and densities of the spheroids, but combined with the
neglected factor, we expect the real numerical errors to become larger
than Juno's error bars. This is confirmed by numerical calculations in the next section.
Therefore, even in this ideal case, the CMS method in its basic form is intrinsically not precise enough to safely fulfill Juno's constraints.
Calculations with 512 MacLaurin spheroids with a 
linear spacing cannot safely enter within Juno's error bars. 

HM16 proposed a better spacing of the spheroids, 
with twice more layers above $50 \%$ of the radius than underneath.
As shown in Appendix \ref{maths_hub},
the new error for 512 spheroids becomes: 

\begin{gather}
|\Delta J_{2}|\times 10^{6} = \dfrac{9}{256} \dfrac{\pi <P_2>}{N^2}\times 10^{6} 
\approx 1.51 \times 10^{-2}.
\label{J2_hub}
\end{gather}

\noindent This is about 2.5 times better than with the linear spacing. 
But once again, when considering the factor of ten neglected in Eq.(\ref{C1_C2}) in nearly the entire 
planet and the strong approximation of perfect density and perfect 
shape of the discrete spheroids, it seems very unlikely to reach in reality a precision
well within Juno's requirements. Again, this is confirmed 
numerically in \S\ref{numerics}.
We note in passing that the difference between the errors obtained with the above two different 
spacings shows the strong impact of the outermost layers in the method.

In conclusion, these analytical calculations show that, at least with 512 sheroids, the CMS method, 
first developed in H13 and improved in 
HM16, leads to a discretization error larger than Juno error bars. Increasing the number of spheroids 
would give a value of $J_2$ outside Juno's error bars and would thus require a change in the derived physical quantities (core mass, 
heavy element mass fraction ...). We investigate in more details the uncertainties on these quantities in \S\ref{ssec : errors_Y}.

It should be mentioned that \cite{WH16} calculated similar errors by comparing their respective methods, finding (their Eq. (15)):
\begin{gather}
\log_{10} |\frac{\Delta J_2}{J_2}| \simeq 0.7 - 1.81 \log_{10}(N),\\
\text{while we get} \\
\log_{10} |\frac{\Delta J_2}{J_2}| \simeq -0.6 - 2 \log_{10}(N).
\end{gather}
The 0.7 constant term is about 20 times the value obtained in Eq.(\ref{J2_hub}). This confirms that, indeed, we underestimated the errors in the analytical calculations, as mentioned above. 

These expressions show that one needs to use at least 2000 spheroids to reach Juno's precision,
as can already be inferred from \cite{WH16}. These authors, however,
suggest that 512 spheroids can still be used to derive interior models, 
because a slight change in the density of one spheroid could balance
the discretization error. A change in one spheroid density, however, will affect the whole
planet density structure, because of the hydrostatic condition. Changing the density of each spheroid
(within the uncertainties of the barotrope itself) can thus yield quite substantial changes in the gravitational moments.

Even more importantly, if the method used to analyze Juno's data has uncertainties due to the number and repartition of 
spheroids larger than Juno's error bars, this raises the question of the need for a better 
precision on the gravitational data, since the derived models depend on the inputs of the method itself. It is thus essential
to quantify precisely the uncertainties arising from the discretization errors, as we do in \S\ref{sec:SCVH} and \ref{sec:BC}.

Another option would be to use Wisdom's method (\cite{Wisdom1996}), which
has virtually an unlimited precision. Table 3 of \cite{WH16} indeed shows that 
its accuracy is greater than anything Juno will ever be able to measure. 
In its present form, however, it is not clear whether this method can
easily handle density, composition or entropy discontinuities;
it is certainly worth exploring this issue.
The CMS method, in contrast, is particularly well adapted to such situations, characteristic of planetary interiors.
As for the standard perturbative theory of Figures, it becomes prohibitively cumbersome when deriving high-order
(fifth even order for J$_{10}$) moments (see e.g., \cite{Nettelmann2017}).
For these reasons, the CMS method appears to be presently the most attractive one 
 to analyze Juno's data. As mentioned above, it is thus
crucial to properly quantify its errors and examine under which 
conditions it can be safely used to derive reliable enough Jupiter models.

\subsection{Spacing as a power of k}

As shown in Appendix \ref{maths_cubic}, 
for a cubic 
distribution of spheroids as a function of depth, 
we obtained the error on $J_2$: 

\begin{equation}
|\Delta J_{2}^{cubic}| \sim \dfrac{\pi}{12} \dfrac{<P_2>}{N^2}.
\label{J_2k_cubic} 
\end{equation}
This is the same result as Eq.(\ref{J_2_fin}).
We checked that we get the same results for a quadratic spacing or with a spacing of power of four 
(it is probably easy to prove that this holds for any integer power of $n$).
This is interesting for several reasons. 

First, these types of spacings have more points in the high atmosphere region,
where the neglected factor of Eq.(\ref{C1_C2}) is closer to two than ten. 
The global neglected factor is then smaller than for a linear spacing.
Second, we studied where the cubic spacing is equal to the linear one. Using Eq.(\ref{delta_ri}), we got:
\begin{gather}
\Delta r_i^{{\rm cubic}} = \Delta r \Longleftrightarrow \dfrac{\Delta i^3}{N^2} = 1 \nonumber \\
\Longleftrightarrow 3i^2+3i +1  =  N^2 \nonumber \\
\Longrightarrow i \sim \dfrac{N}{ \sqrt3}.
\end{gather}

\noindent More generally, with a spacing of power $k$, $\Delta r_i^k$ is smaller than
the linear $\Delta r$ for $i \le {N}/{k^{\frac{1}{k-1}}}$, and larger for larger values of $i$.
That means that above $r_i = a_J \left(1 - 1/k^{\frac{1}{k-1}} \right)$
the radius, potential and density of the spheroids are better estimated than in 
the linear case and less well below this radius. And the stronger the exponent in the spheroid distribution, the larger the external 
gain (internal loss) on the spheroids because the more (less) tight the external
(internal) layers. 

With a quadratic spacing, the spheroids are closer than in the linear
case above $75 \%$ of the radius of 
the planet while inside this limit the spacing is worse than in the linear case. With a cubic spacing the precision in the spheroids repartition is better than in the linear
case down to only 
$80 \%$ of the radius, while
for a power of 4 the limit is $85 \%$. 
%Given the density profile of the gravitational moments 
%(see Figure (\ref{impact_J2})), the quadratic spacing appears to be the best 
%compromise between external gain and internal loss on the spheroid distribution.
As shown and explained in \S\ref{Junoerror}, the cubic spacing ends up being the best compromise.

There is one remaining problem with a cubic spacing: for $N>512$, the sizes
of the first (outermost) layers are extremely small, well below the kilometer,
with densities smaller than $10^{-3}$ kg m$^{-3}$.
It is thus mandatory to have an eos accurate enough in this regime; if 
not the errors will be very large. The other solution, that is
discussed in \S\ref{sec:BC} is to impose a non zero 
density as outer boundary condition.
%In that matter, we also propose an 
%exponential repartition of the spheroids.

\subsection{Exponential spacing of the layers}
\begin{figure}[ht!]
% \captionsetup{justification=centering}
\resizebox{\hsize}{!}{\begin{subfigure}{.5\textwidth}
  \centering
  \includegraphics[width=1\linewidth]{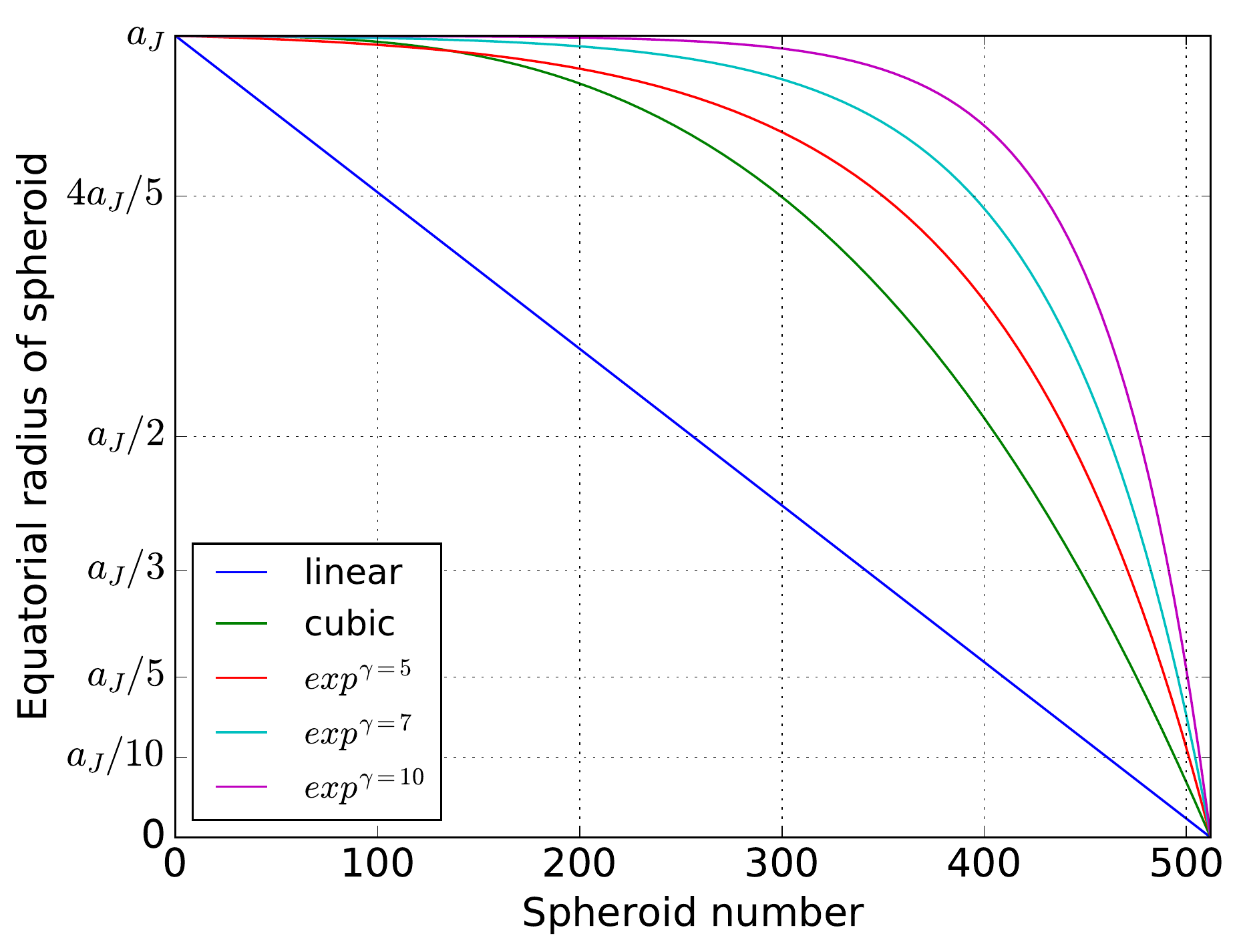}
  \caption{}
  \label{repartition}
\end{subfigure}}
\\
\resizebox{\hsize}{!}{\begin{subfigure}{.5\textwidth}
  \flushleft
  \includegraphics[width=1\linewidth]{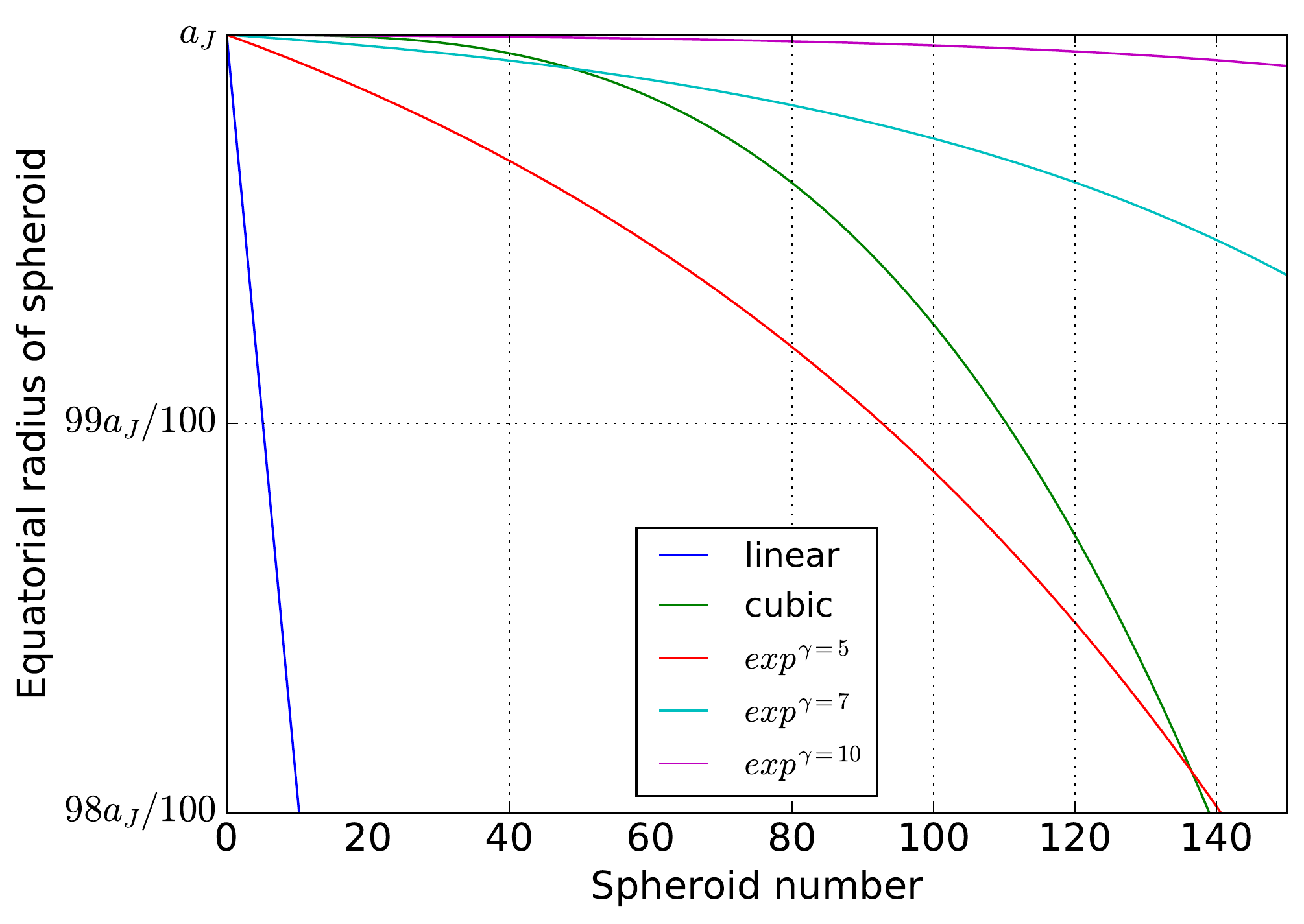}
  \caption{}
  \label{repartition_zoom}
\end{subfigure}}
\caption{Equatorial radius as a function of the spheroid number for various functional forms for the repartition of the spheroids. 
(a): from 0 to $a_J$;
(b): zoom of the external layers.}
\label{fig:repartition}
\end{figure}

Here, we examine an exponential repartition of spheroids.
The details of the calculations are given in Appendix \ref{maths_exp}, and
the repartition functions of the spheroids are displayed in Figure \ref{fig:repartition}. 

\noindent For the
exponential repartition, the value of the error is:

\begin{equation}
|\Delta J_{2}^{exp}| \times 10^{6} \simeq \dfrac{\pi <P_2>}{252}\times \dfrac{\gamma^2}{ N^2}\times 10^{6}
\approx 449 \left(\dfrac{\gamma}{N}\right)^2,
\label{J2_exp}
\end{equation}
where $\gamma \in [5,10]$  is a parameter which determines how sharp
we want the exponential function to be. A large
value of $\gamma$ leads to a steep slope in the innermost layers and a
very small $\Delta r_i$ in the exterior. 
With $N=512$: 
\begin{equation}
|\Delta J_{2}^{exp}|\times 10^{6}  \in [1.54\times 10^{-2}, 1.6 \times 10^1].
\end{equation}
As seen, the smaller $\gamma$ the smaller the error, but also 
the larger the range of outer layers which is a problem, as examined
in \S\ref{numerics}. 
Note also that the neglected factor of
Eq.(\ref{C1_C2}) in the analytical calculations is much smaller because of the very large number of spheroids
in the outermost part of the envelope (see Figure \ref{fig:repartition}).
Compared to Eq.(\ref{J2_hub}), we see that the exponential error is always larger for
$\gamma > 3$. As discussed in the next section, however, the impact of the first layers
becomes very important with this type of spheroid repartition.\\

In summary, in this section, we have shown analytically that 512 layers spaced linearly is not enough to
obtain sufficient accuracy on the gravitational moments of Jupiter to safely exploit Juno's data. 
Indeed, the theoretical errors are within Juno's error bars but
the simplifying assumptions used in the calculations (in particular perfect shape and density of the spheroids) 
suggest that the real calculations
will not match the desired accuracy. As explained in \S \ref{sec:intro},
the method, mathematically speaking, is precise up to $10^{-13}$ for a piecewise density profile,
but the errors due to the discretization, as derived in this section, are of the order of $\sim 10^{-7}$.
Using a better repartition of spheroids yields, at best, 
an error of the order of Juno's error bars but, in any case,
remains insufficient with 512 spheroids. The impact of 
such an error on Jupiter's derived physical quantities is discussed in \S\ref{ssec : errors_Y}.

According to this analytical study, using a minimum of 1000 spheroids spaced 
cubically, quadratically or exponentially seems to provide satisfying solutions. That may sound surprising since the 
spacing used in HM16, with their particular first layer density profile, gives a smaller uncertainty in Eq.(\ref{J2_hub}) than
the cubic or square one in Eq.(\ref{J_2k_cubic}). 
To explain this apparent contradiction, we need to explore more precisely how the error
depends on the different spacings by comparing numerical calculations
with the analytical predictions. This is done in the next section.

\section{Numerical calculations}
\label{numerics}
\subsection{How to match Juno's error bars}
\label{Junoerror}

In this section, we explore two main issues: how does the 
error depend (i) on the number of layers and (ii) on the spacing
of the spheroids. Another parameter we 
did not mention in the previous section is the repartition of the
first layers. This is the subject of the next section. 

To study this numerically, we have developed a code similar to the one developed by Hubbard,
as described in H13. 
The accuracy of our code has been assessed by comparing our results with all the ones published in the above papers.
In the case of constant density,
we recovered a MacLaurin spheroid
up to the numerical precision, while 
for a polytrope of index $n=1$, we recovered the
results of Table 5 of H13, as shown 
in Table \ref{diff_hub2013} (the differences are due to the 
fact that the published results were not converged to the machine precision. With the very same
conditions, our code and Hubbard's agree to $10^{-14}$). We have also performed several tests
of numerical convergence with different repartitions of spheroids.
We have compared our various numerical evaluations of the errors, double 
precision with 30 orders of gravitational moments and 48 points of Gauss-Legendre
quadrature, to a quadruple precision method with 60 orders
of moments and 70 quadrature points, from a couple of hundred
spheroids to 2000 spheroids. The differences are 
of the order of $10^{-13}$, negligible compared with
Juno's error bars. Finally, as 
explained in \cite{Hub2014}, we have implemented an auto exit of the program
if the potential diverges from the audit points.

\begin{table}[ht!]
\begin{center}
\caption{Difference between the values calculated with our code and the results from \cite{Hub2013}.}
\label{diff_hub2013}
\begin{tabular}{cccc}
\hline
\hline
 Quantity & Theoretical value & H13 & Our Code \\
\hline
$J_2 \times 10^{6}$ & 13988.511 & 13989.253 & 13989.239 \\
$-J_4 \times 10^{6}$ & 531.82810 & 531.87997 & 531.87912\\
$J_6 \times 10^{6}$ & 30.118323 &   30.122356 & 30.122298\\
$-J_8 \times 10^{6}$ & 2.1321157 & 2.1324628 & 2.1324581 \\
\hline
\end{tabular}
\end{center}
\end{table}

\begin{figure}[ht!]
\centering
% \captionsetup{justification=centering}Figure
\resizebox{\hsize}{!}{\includegraphics[width=\linewidth]{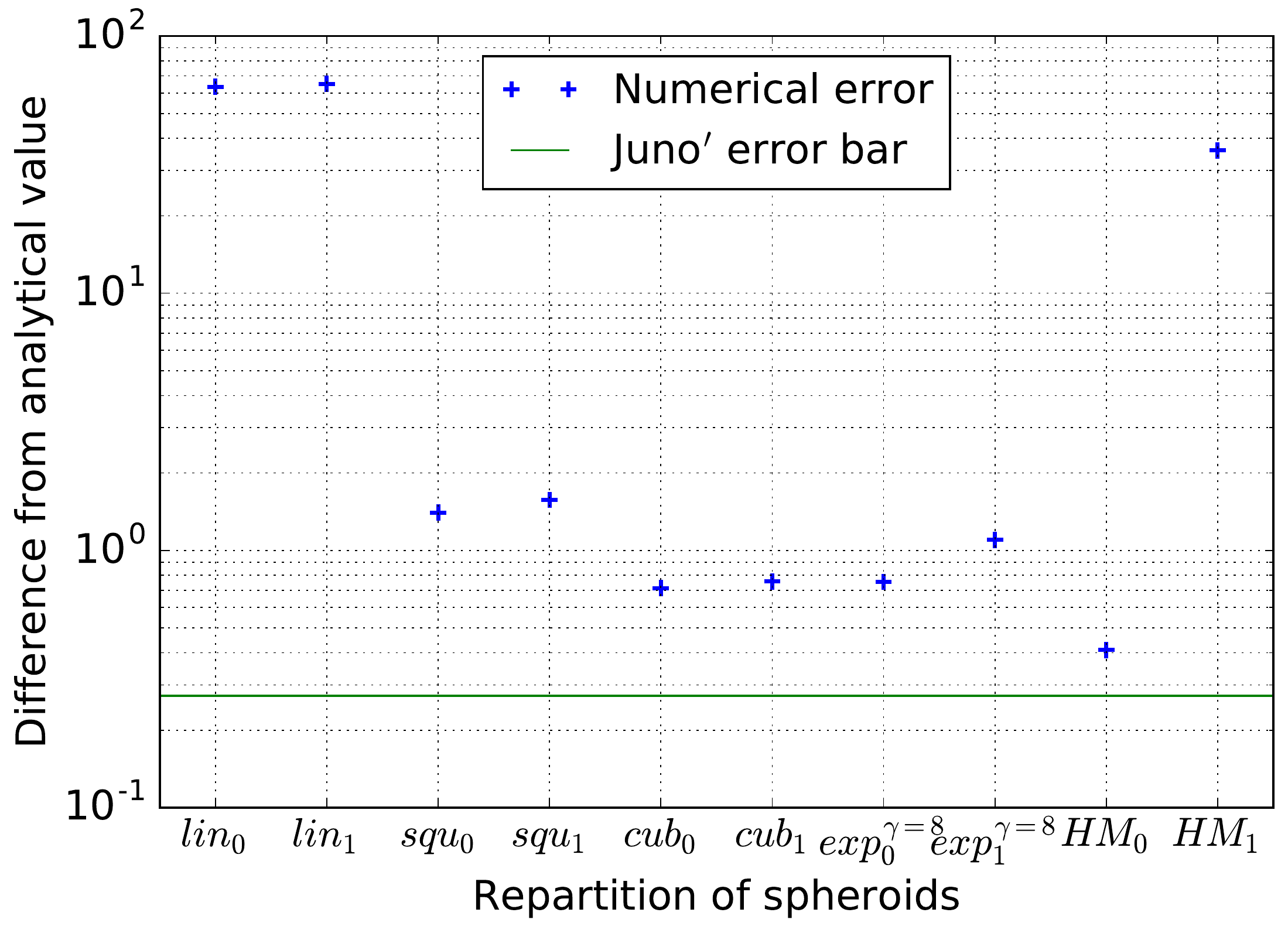}}
\caption{Absolute value of the difference between the expected and
numerical results on $J_2 \times 10^6$ for different spacings of the spheroids with N=512. 
Lin, squ and cub are repartitions of spheroids following a power law
of exponent 1, 2 or 3, respectively; $exp^{\gamma=8}$ is the
exponential repartition with our favored value $\gamma=8$.
HM is the method exposed in HM16, 
explained in Appendix \ref{maths_hub}. 
The subscript $1$ means that the first layer has a non-0 density while the subscript
$0$ means zero density in the first layer.}
\label{diff_512}
\end{figure}

Figure \ref{diff_512} displays the difference between the expected and numerical values of $J_2$ for 
a polytrope of index $n=1$, for different spacings with 512 layers.
The first thing to note is that, whatever the repartition, the numerical errors are always larger than 
Juno's error bars: 512 spheroids are definitely not enough to match the measurements of the Juno mission with sufficient accuracy, 
as anticipated from the analytical calculations of the previous section. 

The second and quite important problem, as briefly aluded to at the end of \S\ref{analytics}, is the difference between calculations using a density $\rho=0$, as in HM16,
 or $\rho\ne 0$
in the first layer. With 512 spheroids spaced as in 
HM16, the first layer is $50$ km deep. Imposing
a zero density in this layer is equivalent to decreasing the radius of the 
planet by $50$ km, which is larger than the uncertainty on Jupiter's radius. 
As seen in the Figure, if one does so, the error on
$J_2$ is more than 100$\times$ Juno's error bar. Specifically,
for the HM16 spacing, when the first layer has a non-zero density, 
its impact on the value of $J_2\times 10^6$ is $\sim 5 \times 10^{1}$.
%Such an error on the first layer is 
%not accepTable, and we don't want to impose an artificial
%layer of zero density to converge our results. 
We discuss that 
in more details in \S\ref{first_layers}.

\begin{figure*}[ht!]
% \captionsetup{justification=centering}
\begin{subfigure}{.5\textwidth}
  \centering
  \includegraphics[width=1\linewidth, height=8cm]{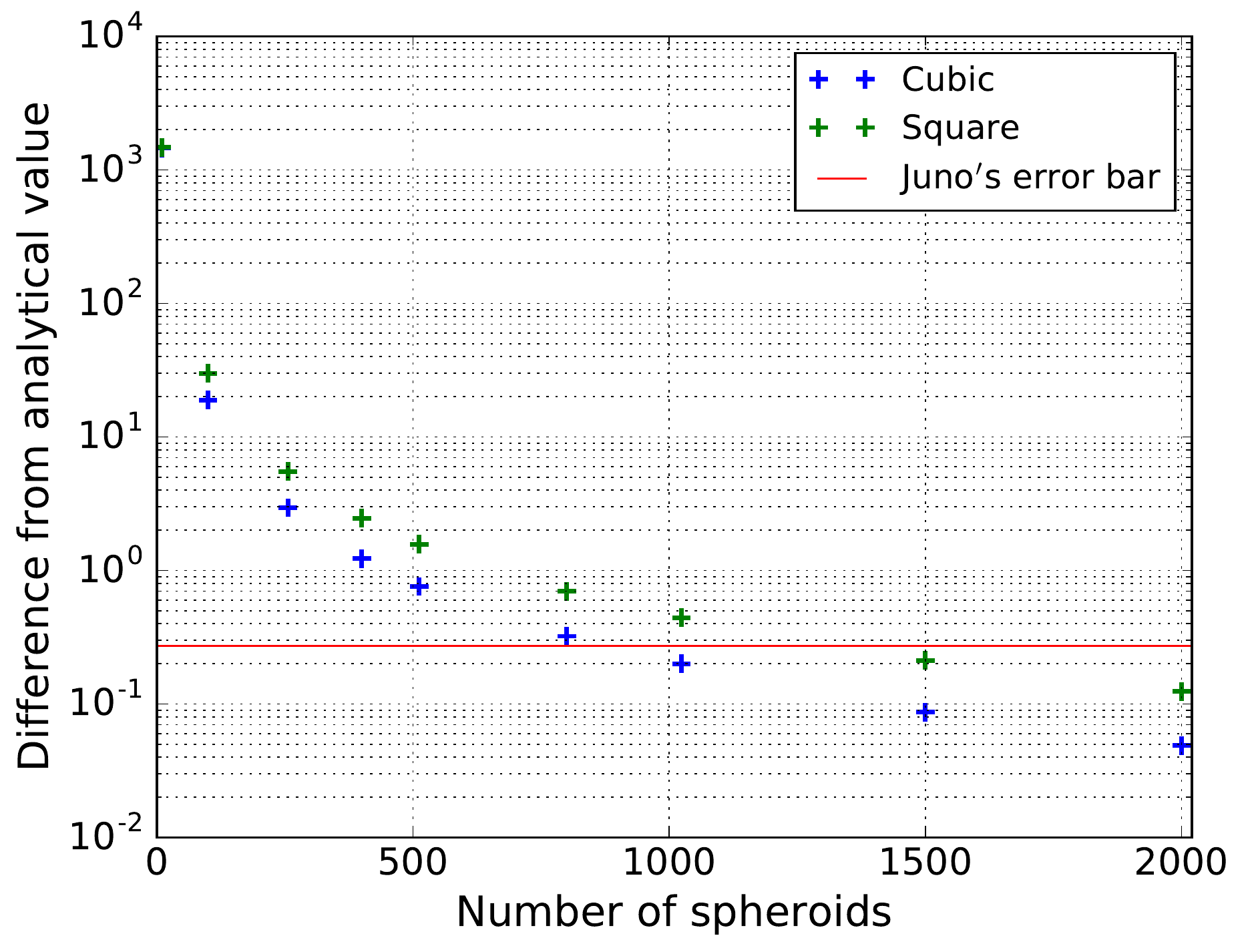}
  \caption{}
  \label{cubic_error}
\end{subfigure}%
\begin{subfigure}{.5\textwidth}
  \flushright
  \includegraphics[width=1\linewidth, height=8cm]{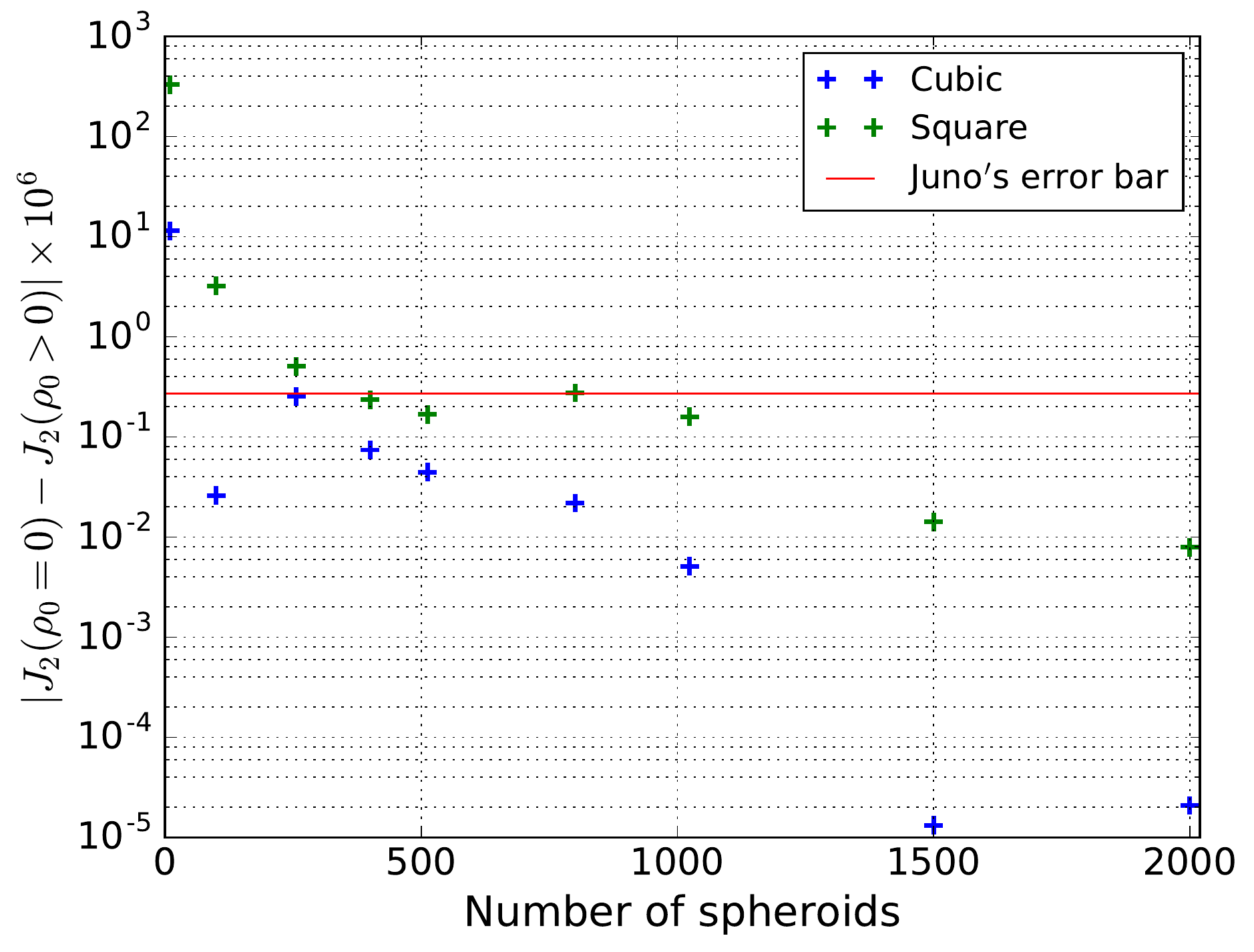}
  \caption{}
  \label{cubic_diff}
\end{subfigure}
\caption{Numerical error as a function of the
number of spheroids for the cubic and 
square repartitions.
(a): error between the analytical and measured $J_2\times 10^6$;
(b) : difference on $J_2\times 10^6$ when the first layer
has 0 or non-0 density.}
\label{cubic_fig}
\end{figure*}

\begin{figure*}[ht!]
% \captionsetup{justification=centering}
\begin{subfigure}{.5\textwidth}
  \centering
  \includegraphics[width=1\linewidth, height=8cm]{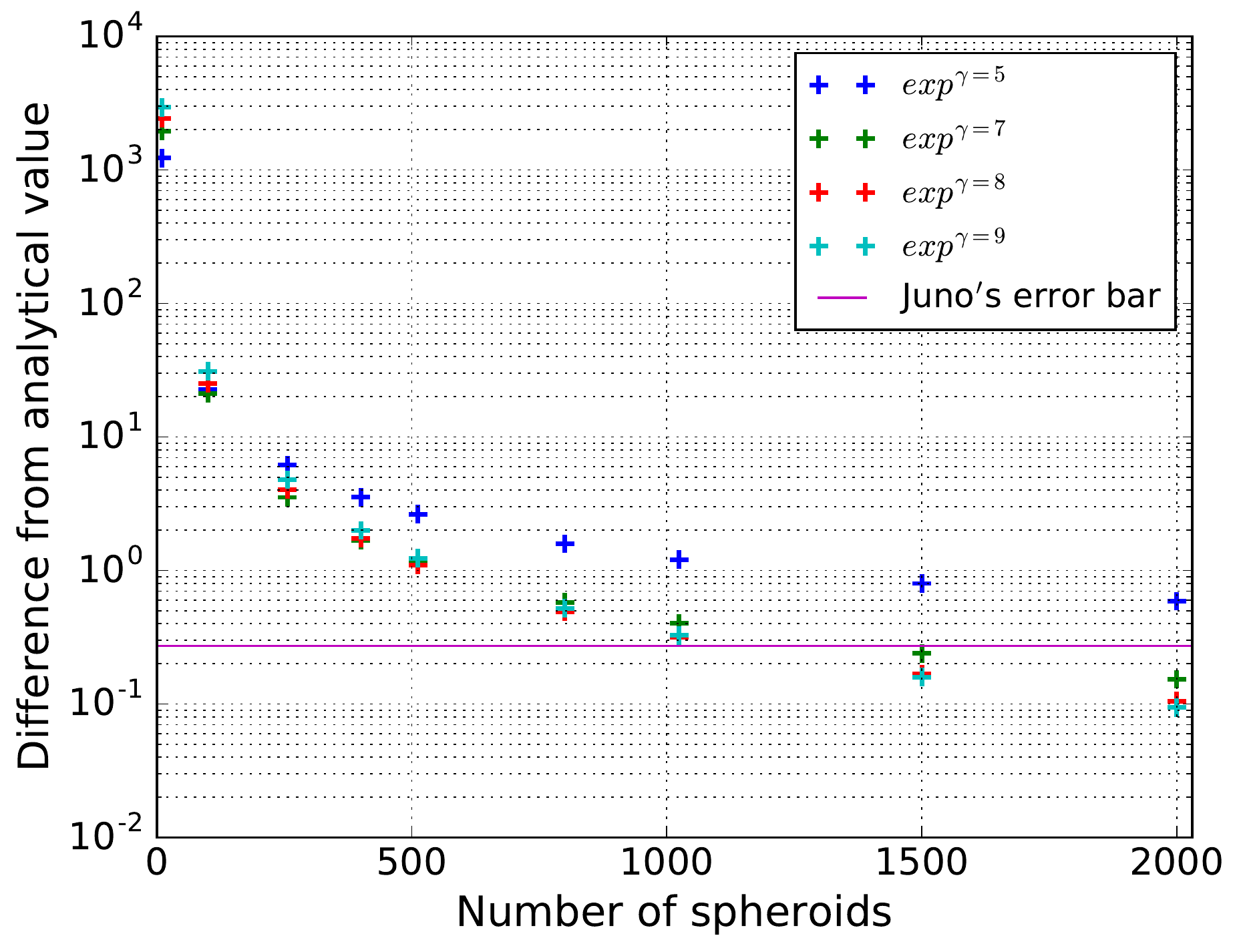}
  \caption{}
  \label{exp_error}
\end{subfigure}%
\begin{subfigure}{.5\textwidth}
  \flushright
  \includegraphics[width=1\linewidth, height=8cm]{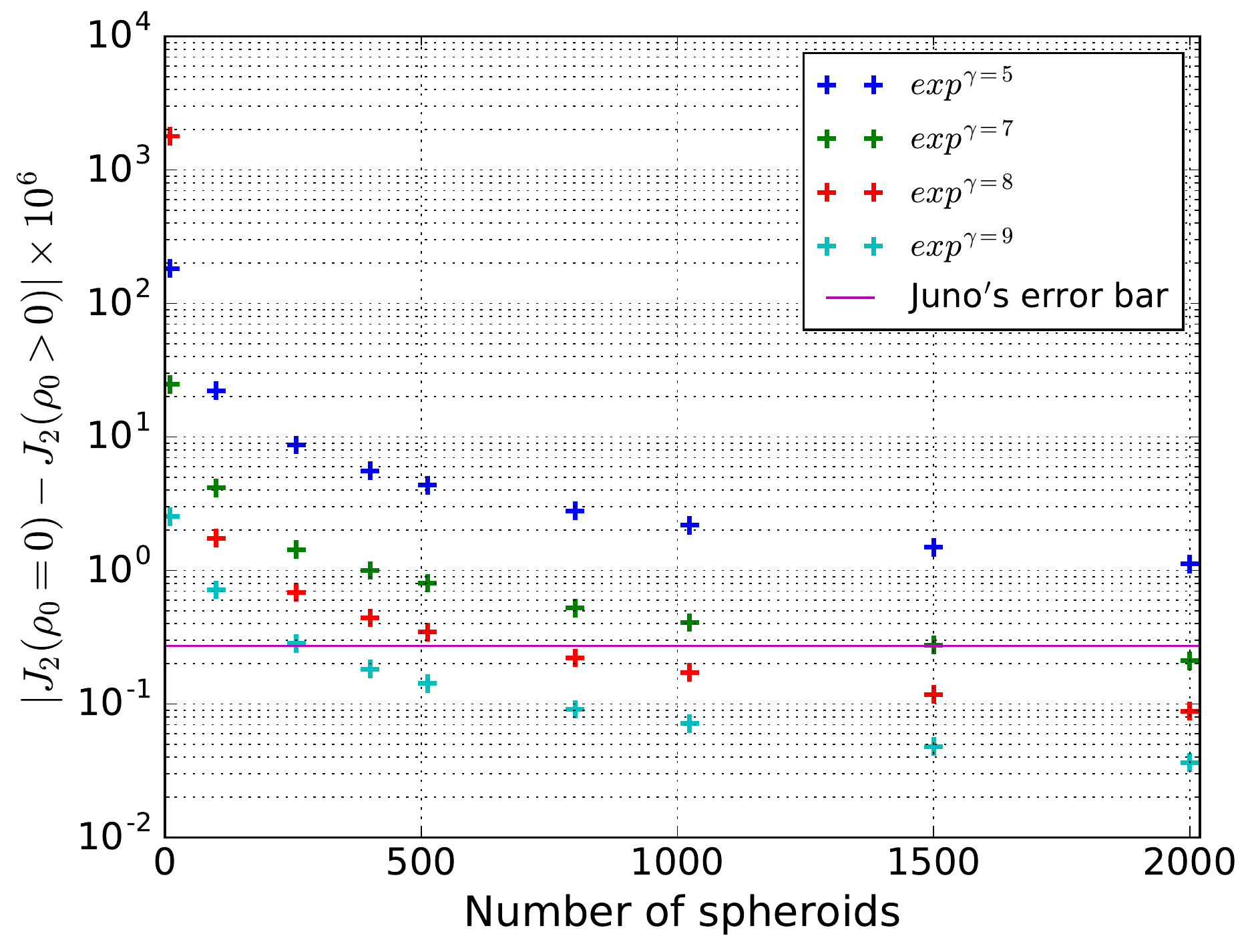}
  \caption{}
  \label{exp_diff}
\end{subfigure}
\caption{Numerical error as a function of the
number of spheroids for exponential repartitions. 
(a): error between the analytical and measured $J_2 \times 10^6$;
(b): difference on the $J_2 \times 10^6$ when the first layer
has 0 or non-0 density.}
\label{exp_fig}
\end{figure*}

\noindent Figures \ref{cubic_fig} and \ref{exp_fig} display
the behavior of the error as a function of the number $N$ of spheroids
for a cubic, square and a few exponential repartitions, respectively. 
As mentioned earlier, we see in Figure \ref{cubic_error} that only for $N>1000$ spheroids do we
get an error smaller than Juno's error bar. 
Importantly enough, we also see that the first spheroid ($N=0$) makes no 
difference on the result when the layers are cubically spaced
(Figure \ref{cubic_diff}). This contrasts with the results obtained with the HM16 spheroid 
repartition, where we see that the first spheroid has a huge impact
on the final result (Figure \ref{diff_512}). The cubic
spacing with at least 1000 spheroids thus matches the constraints, with both 
an uncertainty within Juno's error bars and a negligible dependence
on the first layer. 
As mentioned in the previous section
(and shown later in Figure $\ref{delta_i_zoom}$), however, with the cubic spacing the 
densities of the first layers are orders of 
magnitude smaller than 10$^{-3}$ kg m$^{-3}$, questioning the validity of the description of the gas
properties by an equilibrium equation of state.
For this reason, it is preferable to use a square or exponential spacing in the case of a zero pressure
outer boundary condition. The effect of changing this boundary condition is discussed in \S\ref{sec:BC}.

As seen in Figure \ref{cubic_error}, the square spacing 
requires at least 1500 spheroids to be under the 
error bars of Juno with reasonable accuracy. 
For 1500 or 2000 spheroids, 
the first layers have densities 
around 10$^{-2}$ kg m$^{-3}$. 
In Figure \ref{cubic_diff}, we confirm that, in that case, having
a zero or non-zero density outer layer changes the value under 
the precision of Juno. 

For the exponential spacing, we recall
that the higher the $\gamma$ value the smaller the first layer,
which can become a problem (see \S\ref{first_layers}).
Considering Figures \ref{exp_error} and \ref{exp_diff}, 
the best choices for $\gamma$ are 
$\gamma =8$ or $\gamma = 7$, because it does not require too many spheroids
to converge and at the same time the obtained values for the
densities of the first layers are comparable to the ones obtained with the square
repartition (slightly larger for $\gamma =7$). 

Globally, we conclude in this section that we need at least 1500
and preferentially 2000  spheroids spaced quadratically or exponentially 
to fulfill our 
goals of both high enough precision for Juno's data and densities high enough to justify the use of an equilibrium eos. Having two 
different repartitions that match Juno error bars 
brings some confidence in results that are obtained with a realistic eos instead of a polytrope. 
Furthermore, with such choices of repartition, we do not need to impose a 
first layer with zero density. In \S\ref{sec:BC}, however, we show
that the cubic repartition can also be used, with a different boundary condition.

\subsection{Importance of the first layers}
\label{first_layers}

As explained in HM16, these authors improved the 
spacing of their spheroids by adding up a first spheroid 
at depth $a_J-\Delta r/2$, instead of $\Delta r$,
which has zero density: $\rho_0= 0$. Although this could seem quite arbitrary,
it has an analytical justification emerging from
the linear dependence of the density with radius (Hubbard, priv. com.).
The obtained result is indeed closer
to the analytical value of \cite{Hub1975}, but 
it has two downsides. First, it decreases arbitrarily the radius by $50$ km (see discussion above).
For example, for a cubic spacing with 512 spheroids, 
decreasing the radius by $50$ km leads to 
$J_2 \times 10^6 = 13961.23 $ with an error 
of $2.7 \times 10^{1}$ according to the analytical calculation, 100$\times$ Juno's uncertainty. 
Furthermore, there
is no guarantee that this repartition will still be valid with a realistic eos, 
with an exponential dependence of the density upon the radius.
%We have no idea
%how a realistic equation of state would affect the CMS
%calculation. 
In order to assess the robustness of the results and of the CMS method, one needs to confirm that different repartitions yield
similar values
and that the results are not only correct for one given spheroid repartition. 

 More generally, a key question is: why are the first layers so important in the CMS method ? Going back to Figure \ref{impact_J2}, 
we see that the region outside $90 \%$ in radius contributes as much to $J_2$
as the region within the inner $50 \%$. The problem is the slope of the moment: it is 
much steeper in the outside region. Consequently, the behavior of a layer in the outermost region
yields a much larger error than the contribution from the inner layers. 
This simply reflects the fact that most of the planet angular momentum is in its outermost part.\\
Another issue is that, in the first layers, the change in
density is represents at least 10\% of the density itself and can even become comparable to this latter. In contrast, in the internal layers, even a large $\Delta r$ yields a relative change 
of density from one spheroid to another less than $1\%$. Then, the error due to the inevitable
mistake on the true shape of the spheroids is much more consequential in the external layers, where
the relative density change is significant.
Another source of error stems
from the fact that the pressure is calculated iteratively from the hydrostatic equation,
starting from the outermost layers. A small error in the first layers will then
propagate and get amplified along the density profile. 

Equation (10) of HM13 gives the explicit expression of $J_2$ for
each layer. With the notation of the paper:  

\begin{gather}
J_i^2 = - \left(\dfrac{3}{5} \right) \dfrac{\delta \rho_i \int
_0^1 d\mu P_{2k}(\mu) \xi_i(\mu)^5}{\sum_{j=0}^{N-1} \delta \rho_j
\int_0^1 d\mu \xi_i(\mu)^3}, \\
\delta \rho_i = \rho_i - \rho_{i+1} \text{  if  } i > 0, \nonumber  \\
\delta \rho_0 = \rho_0, \\
\xi_i(\mu) = r_i(\mu)/a_J.
\end{gather}

\begin{figure}[ht!]
% \captionsetup{justification=centering}
\begin{subfigure}{.5\textwidth}
  \centering
  \includegraphics[width=1\linewidth]{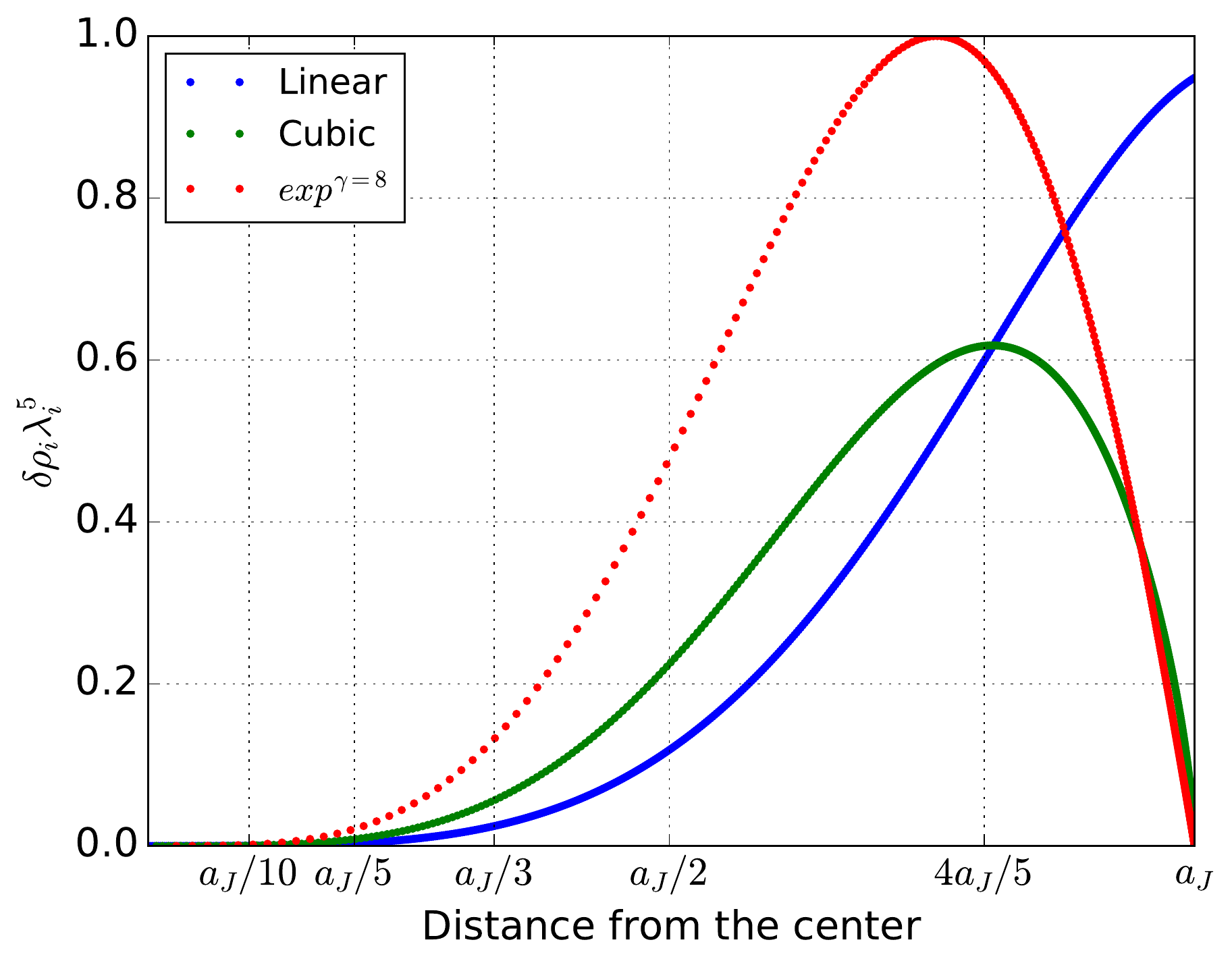}
  \caption{}
\end{subfigure} \\
% \begin{subfigure}{.3\textwidth}
%   \centering
%   \includegraphics[width=1\linewidth]{delta_i_log.pdf}
%   \caption{}
% \end{subfigure} 
\centering 
\begin{subfigure}{.5\textwidth}
  \centering
  \includegraphics[width=1\linewidth]{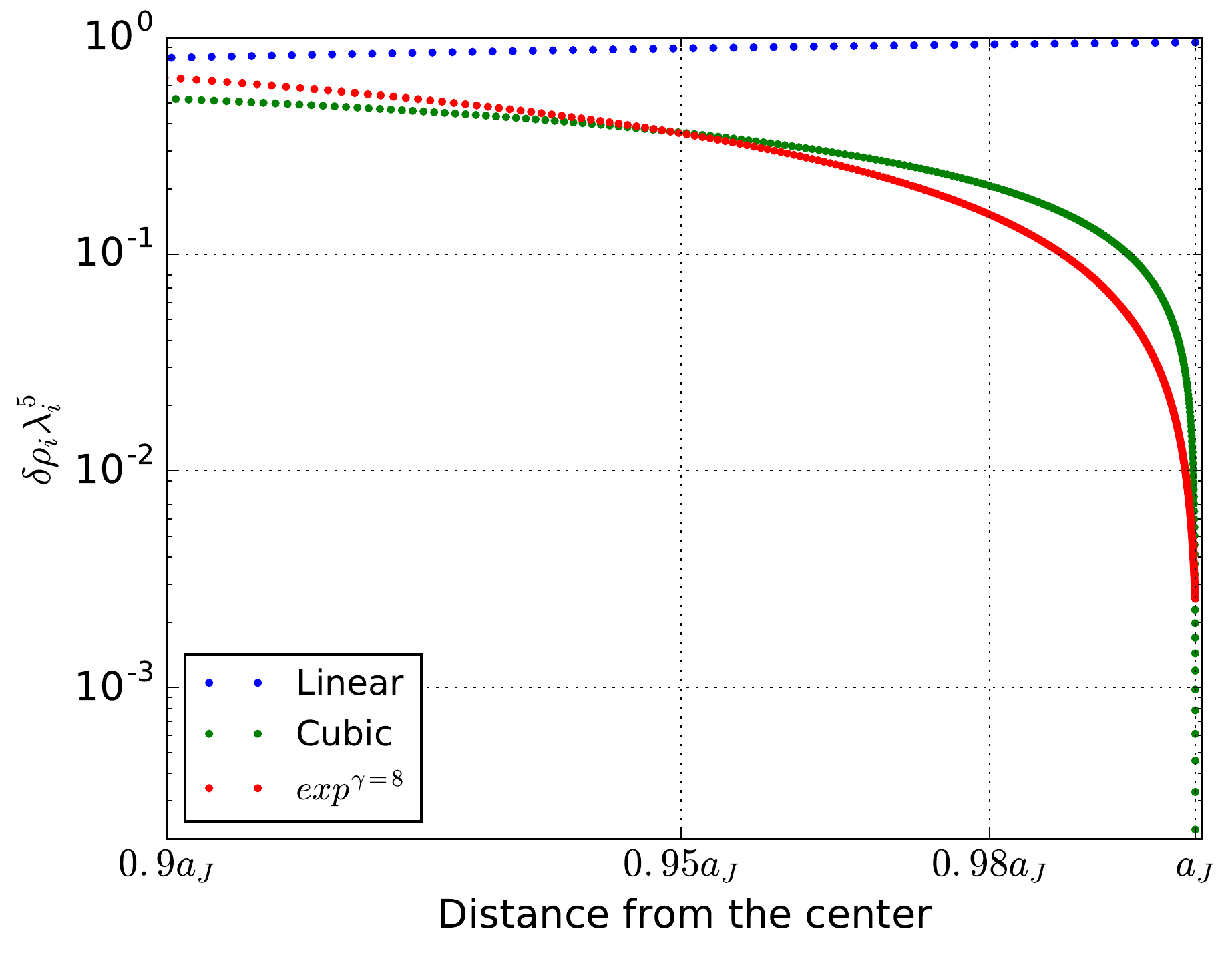}
  \caption{}
  \label{delta_i_zoom}
\end{subfigure}
\caption{Normalized value of $\delta \rho_i \lambda_i^5$ as a function of 
altitude for linear, cubic and exponential ($\gamma =8$)
repartitions of spheroids.
(a) From center to exterior;
% (b) Same in log scale
(b) zoom on the external layers }
\label{delta_i}
\end{figure}

\noindent To illustrate our argument, we said that every spheroid has the same shape
and that $\delta \rho_i$ is the same as in the rest case, given by a
slight modification of Eq.(63) of HM13: 

\begin{gather}
\delta \rho_i = \rho_c \times \delta \lambda_i \left(
\dfrac{\cos(\pi \lambda_i)}{\lambda_i} - \dfrac{\sin{\pi \lambda_i}}{\lambda_i^2} \right), \\
\lambda_i = \xi_i(0) \text{  and  } \delta \lambda_i = \lambda_i - \lambda_{i+1}.
\end{gather}
Then, the only different terms for each layer stemmed from the denominator: 
$\delta \rho_i \times \lambda_i^5$. We plot this value for different spacings of the
spheroids in Figure \ref{delta_i}.

As seen,
the linear spacing is much less precise than the other ones in the
external layers because of 
the very small number of layers; it
will thus greatly enhance the numerical errors. The cubic spacing
is much better but has so many points in the outermost layers that it is
oversampling the extremely high part of the atmosphere. The exponential spacing is the 
worst one in the lower part of the envelope (see Figure \ref{delta_i}), 
but provides a nice cut-off between linear and 
cubic spacings in the uppermost layer region. 

An appealing solution would be to take a combination of all these 
spacings to always have the most optimized one in each region of the 
planet. We have tried this, with a great number
of spheroids, and the conclusion remains the same: the limiting effect
on the total errors is the precision on the external layers
(above $\sim 95 \%$ of the radius). Having
simply an exponential repartition of spheroids throughout the whole planet 
or the same exponential in the high
atmosphere and then a cubic repartition when this latter becomes more precise, and then square and linear repartitions
(such a repartition almost doubles the number of spheroids)
yield eventually the same error.
To obtain results within Juno's precision (see Figures 
\ref{cubic_fig} and \ref{exp_fig}), we need the external layers
to be smaller than 1 km.

To conclude this section, we want to stress that the main
limiting factor in the CMS method is the number of 
spheroids in the outmermost layers (above $\sim 0.95 \times a_J$)
and the proper evaluation 
of their densities. If we do not want to impose
a first layer with 0 density, with consequences not predictable 
in general cases, we must be able to describe this region
of the planet with sufficiently high accuracy. This requires an optimized trade-off
between the number of spheroids in the 
external layers and the value of the smallest
densities. As shown in \S\ref{ssec : errors_Y}, 
the treatment of these layers in the CMS method is the largest source of error in the derived
physical quantities of the planet. 

One could argue that if the outermost layers can not be adequately described by an eos,
a direct measurement of their densities would solve
this issue. Unfortunately, beside the fact that, by construction, Mac Laurin spheroids imply constant density layers, there is an additionnal theoretical challenge: in the 
high atmosphere, the hydrostatic balance barely holds. The isopotential surfaces are 
not necessarily in hydrostatic equilibrium, as differential flows are dominating. 
The time variability in these regions is an other problematic issue. At any rate, the treatment of the
high atmospheric layers is of crucial importance in the CMS method, because it determines the 
structure of all the other spheroids. An error in the description of these layers will propagate and accumulate when summing up the spheroids inward, yielding eventually to large errors. In Section \ref{sec:BC}, 
we examine the solution used in \cite{Wahl17} 
to overcome this issue.

\subsection{1 bar radius and external radius}
\label{1bar}
So far, Jupiter outer radius in the calculations has been
defined as
the observed equatorial radius at 1 bar, $R_{1{\rm bar}}=71492$ km (see Table 1). A question then arises: 
does the atmosphere above 1 bar have any influence on the value of 
the gravitational moments ? 

When converging numerically the radius at 1 bar to the observed value, 
we obtained a value for the outer radius $R_{ext}=71505$ km.
Integrating Eq.(\ref{def_J2k}) from $R_{1{\rm bar}}$ to $R_{ext}$
by considering a constant density equal to the 1 bar density, with the simplifications of Appendix \ref{mu_legendre},
yields a  contribution of the high atmosphere $J_{2_{<1bar}} \times 10^6 \simeq 1 \times 10^{-2}$.

It thus seems reasonable to use the 1 bar radius as the outer radius with the current
Juno's error bars. 
In \S\ref{sec:BC}, however, we examine in more detail this issue and quantify the impact of neglecting the
high atmosphere layers in the calculations, as done in \cite{Wahl17}. In the next section, we now examine if the conclusions of \S\ref{numerics} still hold when using a realistic eos.

\section{Calculations with a realistic equation of state}
\label{sec:SCVH}

The aim of this paper is not to derive the most accurate Jupiter models or to use the most accurate eos,
but to verify (i) under which conditions is the CMS method appropriate in Juno's context, (ii) which spheroid repartition, among the ones tested in the previous sections,
 is reliable
when using a realistic eos to define the barotropes, and (iii) whether the high atmosphere is correctly sampled to induce negligible errors on the moment calculations. 
For that purpose, we have carried out calculations with the Saumon, Chabrier, Van Horn eos (1995, SCvH). This eos perfectly recovers the atomic-molecular perfect gas eos, which is characteristic of the outermost layers of the planet
(\cite{Scvh}). 

%As explained above we have three main issues : is the high 
%atmosphere precisely sampled, is the equation of state valid in the
%region of low pressure and do our conclusions about the spacing 
%stand with a realistic equation of state ? \\

\subsection{Impact of the high atmospheric layers on the CMS method}
\label{ssec:Scvh_high}
\begin{figure}[ht!]
\centering
\resizebox{\hsize}{!}{\includegraphics[width=\linewidth]{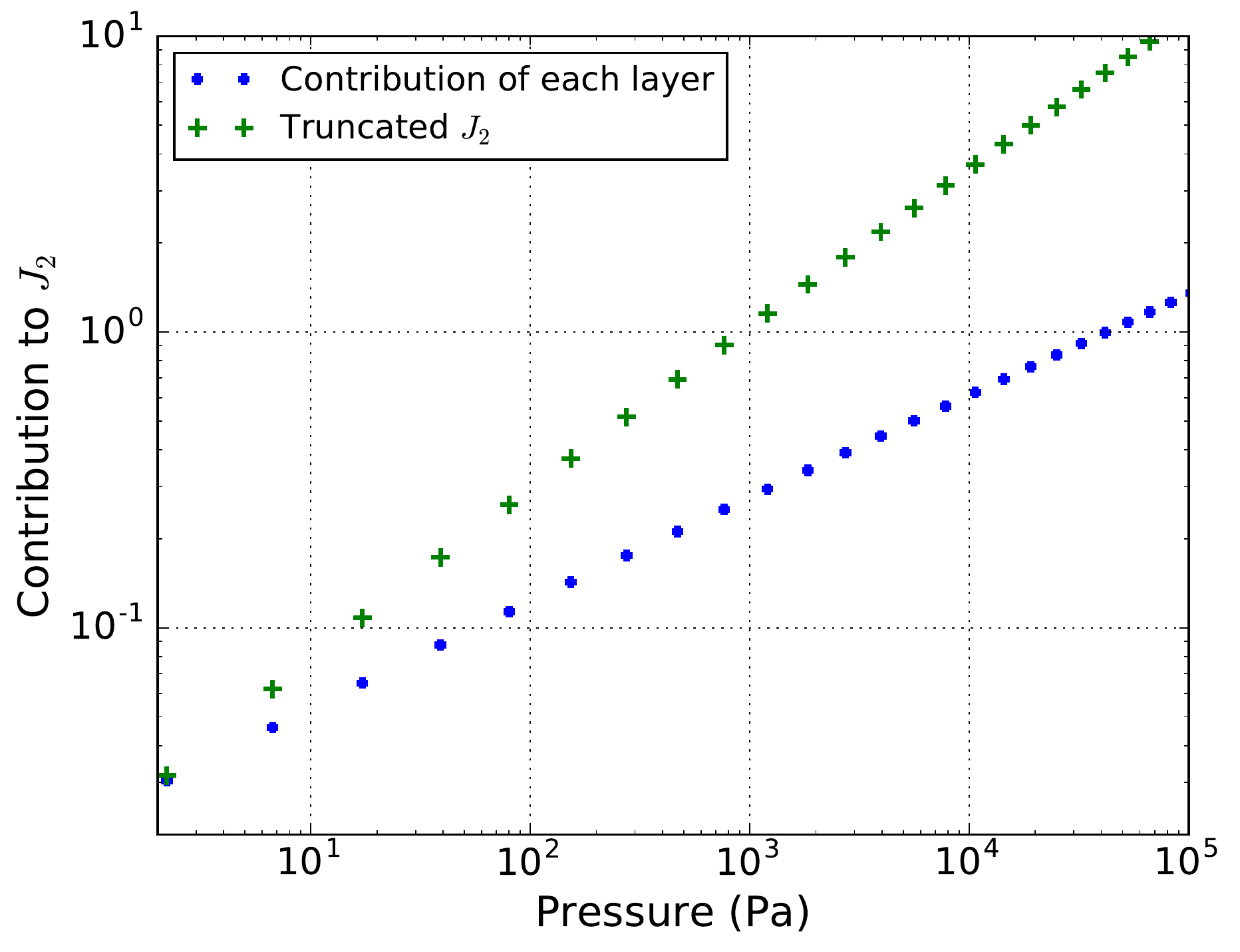}}
\caption{Contribution of each spheroid to the total value of $J_2 \times 10^6$ and to the truncated $J_2\times 10^6$, 
i.e., the sum of the contributions of the 
layers 
from $1$ to $10^{-5}$ bar,
for a square spacing
with the SCvH eos
with 1500 spheroids.}
\label{J2_Scvh}
\end{figure}

Concerning the last of the three aforementioned questions, the SCvH eos
yields much better results than the polytrope. 
The polytropic densities in the high atmosphere are much larger than the realistic ones, yielding an overestimation of
 the contribution of the external layers. Figure \ref{J2_Scvh} illustrates
the contribution to $J_2$ of the outermost spheroids with the SCvH eos. 
Even though, as mentioned earlier, the very concept of an equilibrium equation of state becomes questionable at such low densities ($P<100$ Pa $=10^{-3}$ bar corresponds to $\rho<10^{-3}$ kg m$^{-3}$), this clearly illustrates our purpose.

Figure \ref{J2_Scvh} shows that, when using a realistic perfect gas eos,
using only 1 spheroid up to $10^{-3}$ bar seems to be adequate to fulfill Juno's constraints.
Unfortunately, the real problem is more complex: if we use only one spheroid
of 8 km (a few $10^{-3}$ bar) as a first spheroid, even though its own influence is indeed 
negligible, as if we divide it into many
spheroids, the total value of $J_2\times 10^6$
is changed by $\sim 1-2$, ten times Juno's error !
The reason is that the whole 
density profile within the planet is modified by the inaccurate evaluation
of the first layers.
Indeed, remember that the CMS method decomposes the planet 
into spheroids of density $\delta \rho$. Therefore, a wrong evaluation of
the first layers is propagated everywhere since the density at one level
is the sum of the densities of all spheroids above that level. 
We verified that
any change in the size of the first layer, from less than 1 km to 35 km
significantly changes the value of $J_2$. 

\noindent We have also tried a solution where we converged the 1 bar radius to the observed $R_{1{\rm bar}}$ value, 
in order to avoid this modification in the structure, but the result still holds:
it is not possible to reduce the global contribution of the first layers,
and the densities of the first layers are too small to be realistically
described by an equilibrium equation of state. 
%It is important to notice that we have tried
%this as well with the polytrope and the results stand : 
Clearly, with the CMS method, it is not 
possible to diminish the size of the first layers
without significantly affecting the gravitational moments.
This, again, is examined in detail in \S\ref{sec:BC}.

Table \ref{tab:failure} gives the results for an exponential repartition
of 1500 spheroids using the SCvH eos, with no core (in that case, 
the size of the first layer is 0.6 km). Clearly, varying the size of
the first layers yields changes of the
$J_2$ value larger than Juno error bars.  
A quadratic repartition yields similar results with differences on $J_2 \times 10^6$ that are 
$<3\times 10^{-1}$.
It is not clear, however, whether the induced errors stem from the 
iterative calculations of the pressure from the hydrostatic equation or from the error on the potential
determination. Therefore, we confirm with a realistic eos the results obtained in Section \ref{first_layers}: a very accurate evaluation of the first layers is mandatory in order to properly converge the CMS method.

\begin{table}[ht!]
\caption{Value of $J_2 \times 10^6$ for an exponential $\gamma = 7$ spacing 
with 1500 spheroids and the SCvH eos with no core. Left column: the outer radius is taken as Jupiter's radius; right column: the radius at 1 bar is taken as Jupiter's radius}
\label{tab:failure}
\begin{center}
\begin{tabular}{c || cc }
\hline
\multirow{2}{*}{First layer(km)}& \multicolumn{2}{c}{$J_2 \times 10^{6}$} \\
\cline{2-3}
& $R_{ext} = $71492km&  $R_{1bar} = $71492km\\
\hline 
\hline
0.6 (unchanged) & 14997.24 & 15046.34 \\
3 &14999.03  & 15046.71\\
6 & 15001.11 & 15047.46 \\
15 & 15012.00 & 15051.85\\ 
35 & 15035.41 & 15061.54\\
\hline
\hline
\end{tabular}
\end{center}

\end{table}

%In that regard, more work is needed to determine 
%how to find which choice gives a physical answer to the 
%evaluation of the gravitational moments. We will choose for the next
%subsection, and without strong conviction, 
%to use the full spacing even though the first layers densities are not
%physically determined by the equation of state. \\

\subsection{Errors arising from a (quasi) linear repartition}

% We now examine in detail whether the
% HM16 spacing, which has been used to infer recent Jupiter models in the context of Juno, is consistent or not with the 
% results from the square and exponential spacings.
In order to assess quantitatively the errors arising from an ill-adapted
repartition of spheroids, we have carried out 3 test calculations.
One without a core, where we
changed the H/He ratio (i.e., the helium mass fraction $Y$) to converge the CMS method 
(as explained in HM16, that means having a factor $\beta=1$ in H13).
A second one where we imposed $Y$ and the mass of the core
but allowed the size of the core to vary (we have verified that 
imposing the mass or the size leads to the same results). And, finally,
one where $Y$ and the mass of the core could change and 
we converged $J_2$ to the observed value. These models
were not intended to be representative of the real Jupiter
but to test the spacings
of the CMS explored in the previous sections. All these tests used the 
observed radius as the outer radius, $R_{eq}=71492$ km.
The results are given in Table \ref{Test_scvh}.

\begin{table*}[ht!]
\caption{Difference between square, exponential $\gamma =7$ 
and HM16 spacing for 3 different tests with 1500 spheroids.}
\label{Test_scvh}
\begin{center}
\begin{tabular}{c || ccc || ccc || ccc }
\hline
\hline
 \multirow{2}{*}{Moment}& \multicolumn{3}{c||}{Test 1 : no core} & \multicolumn{3}{c||}{Test 2 : Y = 0.37, $M_c$= 4$ M_T$} &
  \multicolumn{3}{c}{Test 3 : fit on $J_2$} \\
  \cline{2-10}
& Square & Exp$^{\gamma=7}$ & HM16 & Square & Exp$^{\gamma=7}$ & 
HM16 & Square & Exp$^{\gamma=7}$ & HM16 \\
\hline
\hline
$J_2 * 10^{6}$ &14997.42 & 14997.24 & 15025.89& 
14636.63 &  14636.47 &14680.96 &
14696.51 & 14696.51 & 14696.51 \\
$-J_4 * 10^{6}$ & 604.21 & 604.19 & 606.49 & 
586.44 & 586.42 & 589.47&
589.21 & 589.21 & 590.08 \\
$J_6 * 10^{6}$ & 35.79 & 35.79 & 35.99 & 
34.62 &34.62 & 34.87&
34.80 & 34.80 & 34.90 \\
\hline
\hline
\end{tabular}
\end{center}
\end{table*}

The difference on $J_2\times 10^6$ between the square and exponential spacings with 1500 spheroids
are $<2 \times 10^{-1}$, as expected from Figures 
\ref{cubic_error} and \ref{exp_error}. 
In contrast, the HM16 method, which is linear in $\Delta r$, though with a different 
spacing in the high and low atmosphere, respectively,
leads to differences 100$\times$
larger. These values were obtained with a HM16 spacing with 1500 spheroids; with 
512 spheroids the differences from square and exponential
are a factor ten larger. 
This demonstrates that 
the square and exponential repartitions remain consistent, confirming the results obtained for
the polytropic case, whereas the HM16 spacing diverges completely. 
As expected, 
the idea of using a first half layer of zero density fails when using a
realistic eos. 

Another proof is the difference,
with the HM16 repartition, between 512 and 1500 spheroids: in the 
polytropic case the difference on $J_2\times 10^6$ was found to be around $2 \times 10^{-1}$. Indeed
the HM16 repartition 
with 512 spheroids led to results very close to the theoretical value. In our first
test case, the value of $J_2\times 10^6$ we obtained
for the HM16 spacing with 512 spheroids is $15075.52$.
When we compare to Table \ref{Test_scvh}, 
the difference is $5 \times10^{1}$: the discretization error on the gravitational moments
with this spacing is way off Juno's precision. Moreover,
we see that the order of magnitude of the error is similar to the one in the
linear case of Figure \ref{diff_512}. Indeed, if one does not use a polytrope, 
the HM16 spacing behaves as badly as a linear spacing, as expected from 
the quasi linear shape of this repartition. 

\begin{figure}[ht!]
% \captionsetup{justification=centering}
\begin{subfigure}{.5\textwidth}
  \centering
  \includegraphics[width=1\linewidth]{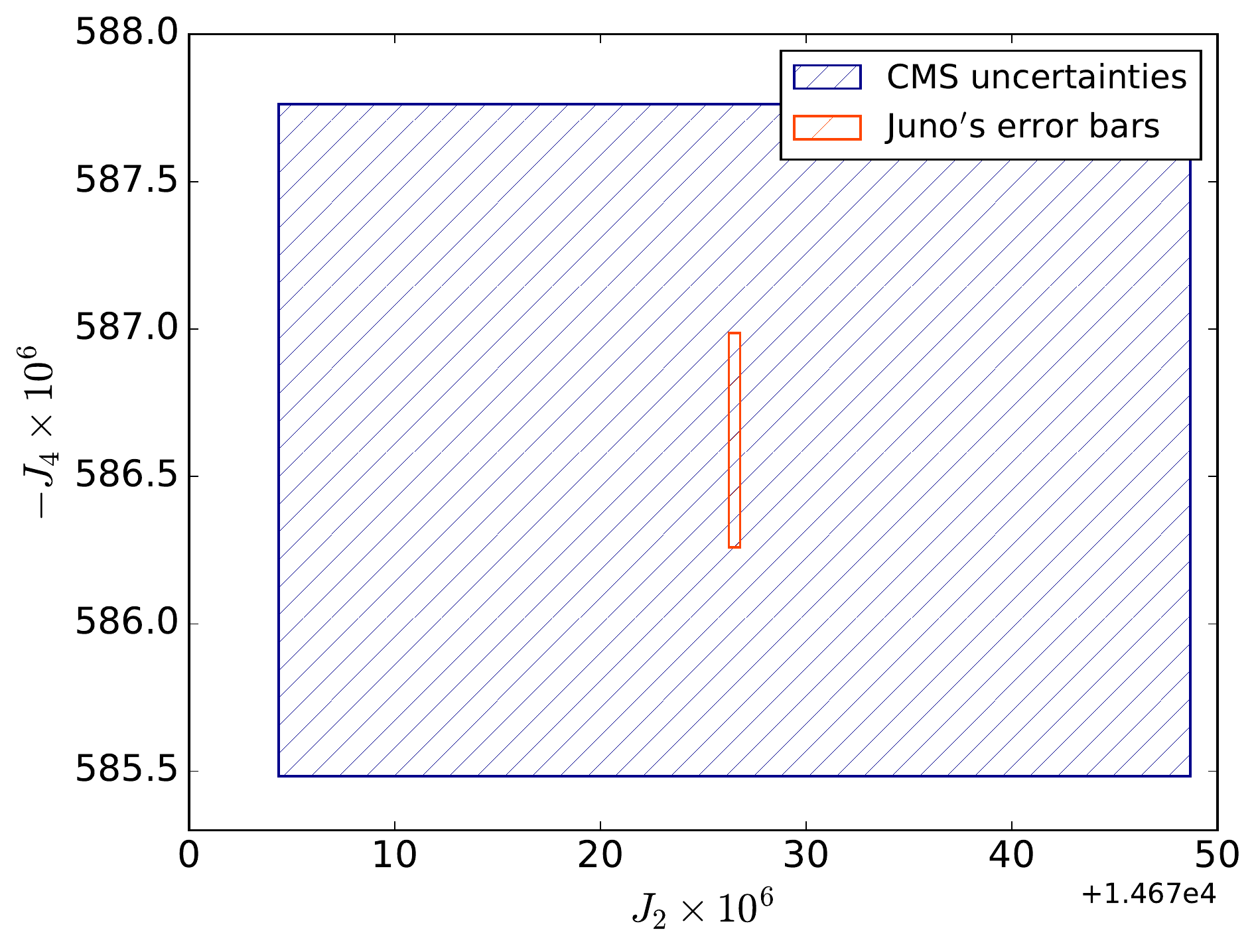}
  \caption{}
  \label{fig:J2J4_Scvh}
\end{subfigure} \\
% \begin{subfigure}{.3\textwidth}
%   \centering
%   \includegraphics[width=1\linewidth]{delta_i_log.pdf}
%   \caption{}
% \end{subfigure} 
\centering 
\begin{subfigure}{.5\textwidth}
  \centering
  \includegraphics[width=1\linewidth]{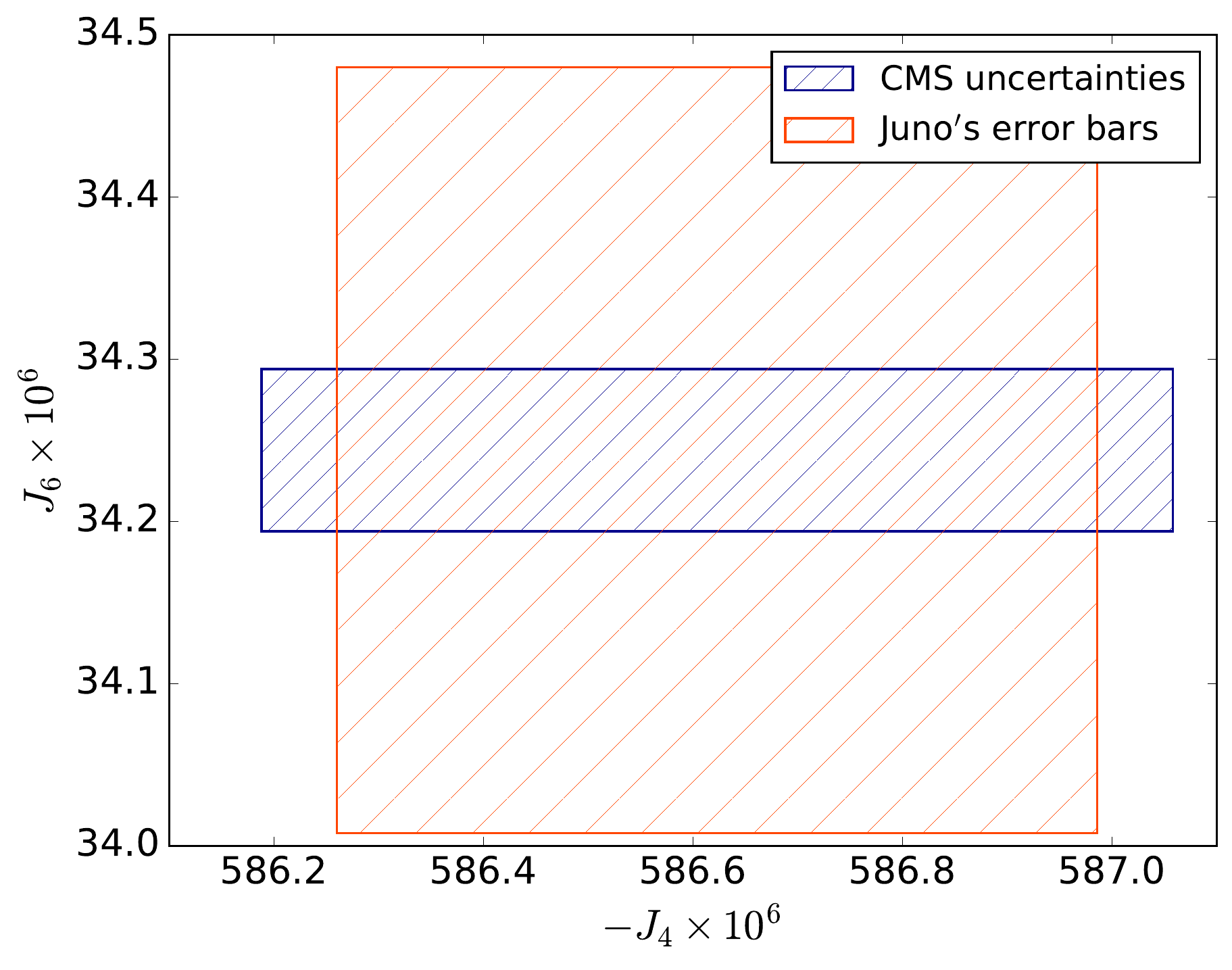}
  \caption{}
  \label{fig:J4J6_Scvh}
\end{subfigure}
\caption{Comparison of the errors of the CMS method from 
quadratic and HM16 repartitions with Juno's error bars, 
shifted to be centered on the observed $J_n$.
(a) $-J_4 \times 10^6$ vs $J_2 \times 10^6$ from test 1;
% (b) Same in log scale
(b) $J_6 \times 10^6$ vs $-J_4 \times 10^6$ from test 3, when $J_2$
is fit. }
\label{fig:Jn_Scvh}
\end{figure}

Figure \ref{fig:Jn_Scvh} summarizes graphically the uncertainties arising from 
using an ill adapted repartition of spheroids. As stated above, 
the errors on $J_2$ from the bad handling of the high atmosphere reach almost
$100 \times$ Juno's error bars, as shown in Figure \ref{fig:J2J4_Scvh}.
On the other hand, when fitting $J_2$, 
Figure \ref{fig:J4J6_Scvh} shows that the errors on $J_4$ and $J_6$ are 
comparable or inferior to Juno's. 
Nonetheless, we show in \S \ref{ssec : errors_Y} that this leads to 
significant changes in the physical quantities, hampering the determination of interior 
models in the context of Juno. These changes can be considered as 
negligible compared to the uncertainties 
on the various physical quantities in the models 
%(core of Jupiter is supposed to weigh between 0 and 
%15 Earth masses, see e.g. \cite{HM2016} or \cite{Leconte2012}) 
but they are significant when one aims at matching
Juno's measurements and thus take full advantage of the Juno mission.

What we have shown so far in the present study is that the CMS method, when taking
into account the entire atmosphere above 1 bar, relies on 
uncontrolled
assumptions that can change significantly the value of $J_2$. The whole 
purpose of the method, however, is to constrain with high precision Jupiter's internal structure (mass fraction of heavy elements, mass of the core ...). Therefore, we need to assess the impact of these
assumptions on such quantities.

\subsection{Intrinsic uncertainties on the Jupiter models}
\label{ssec : errors_Y}

In this section, we derived interior models that match
both $J_2$ and $J_4$ from the first Juno data (\cite{bolton2017}) 
and determined how they were affected by the number of spheroids and their repartition.
Ideally, one would like the final models not to depend on the method used to derive them.
We recall that the purpose of this paper is not to derive the most accurate Jupiter models but to determine the impact of the uncertainties in the CMS method upon these models, in the context of the Juno mission.
Therefore, we used the SCvH eos with an effective 
helium mass fraction (Y$^\prime$) which includes the metal contribution (see \cite{Chab92}). We supposed that
the core was a spheroid of constant density $\rho_{core} = 20000$ kg m$^{-3}$, the typical density of silicates at the pressure of the center of Jupiter.
At 50 GPa, we imposed a change in the 
effective Y$^\prime$ at constant 
temperature and pressure, thus a change in entropy, to mimic either an abrupt metallization of hydrogen or a H/He phase separation. Inward of this pressure, the
value of the effective helium mass fraction is denominated $Y^\prime_2$.  All the models were converged to the appropriate $J_2$ and $J_4$ values with a 
precision of $10^{-8}$, within Juno's error bars.

%In order to converge the models to the appropriate $J_2$ and $J_4$ values (with a 
%precision of $10^{-8}$, within Juno's error bars) we are combining several
%Newton-Raphson schemes. First, we change the value of $Y$ and $Y_2$ by the
%same amount to converge the $\beta$ of H13 (which is equivalent to converge the mass).
%This means changing the global content of metal of the planet, as explained in HM16.
%Then we change the size of the core to match with $J_2$, and finally we alter
%the value of $Y_2$ alone, thus the inner entropy, to match with $J_4$.

Table \ref{Test_Y_mc} gives the values of $Y^\prime$, $Y^\prime_2$ and the core mass for different repartitions
and numbers of spheroids. We considered a quadratic repartition with an unchanged
first layer or with a first layer of 10 km, and the HM16 repartition. With HM16,
the change on the effective helium mass fraction is of the order of 
5\%, consistent with the expectations from the results of the previous
sections. This confirms that the HM16 repartition of spheroids is not suitable to exploit Juno's data. When we used
the $Y^\prime$ and core mass fraction obtained with 512 spheroids to a case with 1500 spheroids, 
the value of $J_2 \times 10^6$ became $14647.9$, which leads to an error of $5 \times 10^1$, a hundred times Juno error bars.
As stated in the 
abstract, this result demonstrates that Jupiter models which have been derived 
with this spheroid repartition are invalid, in the framework of Juno's constraints. The most recent calculations of \cite{Wahl17} are examined specifically in \S\ref{sec:BC}.

\begin{table}[ht!]
\caption{Effective helium mass fraction above (Y$^\prime$) and underneath ($Y^\prime_2$) the entropy 
jump, and mass of the core ($M_c$ in units of $10^{24}kg$) for different repartitions
and numbers of spheroids. Quadratic is a power law spacing of exponent two, Quadratic 10
is the same with a first layer of 10 km and HM16 corresponds to the repartition used in \cite{HM2016}}
\label{Test_Y_mc}
\begin{center}
\begin{tabular}{c | c || c  c  c  }
\hline
\hline
N & & Quadratic & Quadratic 10 & HM16 \\
\hline
\cline{3-5}
 \multirow{3}{*}{512} & Y$^\prime$ & 0.34622
 & 0.34451 & 0.32297\\
 \cline{2-5}
  & Y$^\prime_2$ &0.37379  & 0.37368 & 0.37215\\
 \cline{2-5}
  & $M_c$  & 9.9241 & 9.8952& 9.6819 \\
\hline
\hline
 \multirow{3}{*}{800} & Y$^\prime$ & 0.34698 & 0.34496 & 0.33168\\
 \cline{2-5}
  & Y$^\prime_2$ & 0.37386& 0.37372&0.37276 \\
 \cline{2-5}
  & $M_c$ & 9.9227&9.8928 & 9.7677\\
\hline
\hline
 \multirow{3}{*}{1500} & Y$^\prime$ & 0.34742 &  0.34506& 0.33840\\
 \cline{2-5}
  & Y$^\prime_2$ & 0.37390& 0.37372& 0.37323\\
 \cline{2-5}
  & $M_c$ & 9.9193& 9.9000& 9.8453\\
\hline
\hline
\end{tabular}
\end{center}
\end{table}

With the quadratic repartition, using the values obtained with
512 spheroids in the 1500 spheroid calculations yields an error on $J_2 \times 10^6$ of $4 \times 10^0$, ten times
Juno's error bars (which confirms the necessity to use
at least 1500 spheroids and this type of repartition). When correctly fitting $J_2$ and $J_4$ with 
this repartition, the change on $Y^\prime$
from 512 to 1500 spheroids is about 0.5 $\%$, about the same as the difference between using a
$<1$ km or a 10 km first layer. 
For the core mass, increasing the number of spheroids is useless as
the difference is negligible compared to the one due to changing the size of the first layer.
Therefore, the uncertainties due to the description of the first layers are the dominant sources of limitation in the determination of Jupiter's physical quantities like $Y^\prime$ or the core mass fraction.

In reality, the limitation is even more drastic. First, with a 10 km first layer, the 
outer density is $\simeq 3 \times 10^{-3} $ kg m$^{-3}$. This is too low to use an equilibrium equation of state.
If we use instead a 20 km first layer, increasing the density to more acceptable values, 
the uncertainties  on $Y^\prime$ are about 1-2$\%$ and  $>5 \%$ on $M_c$ .
%More importantly, we already know that the degeneracy is huge between the gravitational moments and the interior models,
Virtually any model leading to such an uncertainty can fit $J_2$ and $J_4$. 
%instead of using a Newton-Raphson scheme on $Y$ and the core mass, we could use one on the first layers and their densities and converge an internal structure that a priori would not fit with the observed $J_2$ and $J_4$. 
Indeed, given the impact of the first layers on the gravitational moments, they could be used, instead of $Y^\prime$ and $M_c$,
as the variables to fit models
%as slightly changing these layers  slightly (for example saying that the first density shouldn't be the density at the  middle of the layer but bigger or smaller) 
with the appropriate gravitational moments. Slightly changing these layers (mean density, size,..) can yield adequate models. Therefore, no model can be 
derived within better than a few percent uncertainties on $Y^\prime$, thus the heavy element mass fraction, and on the core mass.

%One could argue that these conclusions rely on very  simplifying assumptions of the model (SCvH eos, fixed core density ...). 
%Moreover, our models don't necessarily match with $J_6$ or just within the error bars. 
% As there are a lot of poorly constrained parameters
% in the internal structure of Jupiter (abundance of metals, pressure of 
% metalization, entropy change, ... ) we consider these uncertainties to be rather conservative limits. 
% %Indeed in our case, changing the metalization pressure allows a change in the $J_6$ but we could also fit $J_6$ by matching with $J_4$ within the error bars of Juno instead of having a precision of $10^{-8}$.
% Models can not be derived to better accuracy (even when fitting $J_6$)
% because of the intrinsic errors of the CMS method in its present form. 
% %In that regard,  only qualitative physically based model can be derived from the method in its present form. 
% As already mentioned in \S\ref{analytics}, we thus show that, in spite of its mathematical elegance, the CMS method is unable to take full advantage of the accuracy of Juno's data.
% This contrasts with the conclusions of \cite{WH16}.
% % that there is no need to change the method to exploit Juno data.

\indent The calculations carried out in this section yield three major conclusions: 
\begin{enumerate}
\item
It is not possible to use a linear repartition, especially with a zero density 
first layer, as
the values of the moments would be changed by 100$\times$ the value of Juno's 
error bars when increasing the number of spheroids.
\item
The results obtained with a square or an exponential repartition with 1500 spheroids yield
consistent errors, no matter the size of the first layer, and can thus be considered as reliable spheroid repartitions.
The size of the first layer, however, remains the major source of uncertainty in the CMS method
when taking into account the high atmosphere above the 1 bar level.
% Note also that, when using an appropriate eos instead of a polytrope to
% calculate the densities in the first layers, the value of $J_2$ changes drastically.
\item
When deriving appropriate internal models of Jupiter, the CMS method such as used
in HM16 can not allow the determination of key
physical quantities such as the amount of heavy elements or the size of the core to better than a few percents. Changing the 
size and/or density distribution of the first layers yields a significant 
change in the values of the gravitational moments. This degeneracy unfortunately hampers the derivation of very precise models.
\end{enumerate}

In the light of these conclusions, the CMS method, such as used so far in all calculations, needs to be improved. To do so,
\cite{Wahl17} propose to neglect the high atmosphere, and define
the outer boundary at the 1 bar level. 
We examine this solution in the next section.\\

\section{Taking the 1 bar level as the outer boundary condition}
\label{sec:BC}

\subsection{Irreducible errors due to the high atmosphere region ($<$1 bar)}

According to their second appendix, \cite{Wahl17}
impose the 1 bar level as the outer boundary condition in their CMS calculations. As shown
with a polytropic eos in \S\ref{first_layers}, the atmosphere above this level,
called here the high atmosphere, has a contribution to $J_2 \times 10^6  \simeq 1 \times 10^{-2}$.
With a realistic eos, converging the 1 bar radius on the observed value
yields a high atmosphere depth of $\sim70$ km but a smaller outer density than the polytrope. The contribution
to $J_2$ is then about the same.

The high atmosphere, however, also alters the evaluation of the potential inside the planet
and thus the shape of the spheroids in the CMS method. In Appendix \ref{app:phi_neg}, we
evaluated the first order neglected potential $\phi_{neg}$ on each spheroid due to the omission of the high
atmosphere. We showed that it is a constant value, which does not depend
on depth or angle:
\begin{gather}
\phi_{neg} \simeq 2.87\times \dfrac{G M}{a_{1bar}} \left(\dfrac{\rho_{ext}}{\bar{\rho_S}} \right)
\left(\dfrac{\Delta a}{a_{1bar}}  \right),
\label{eq:phi_neg}
\end{gather}
where $\rho_{ext}$ is the (constant) density of the high atmosphere, $\bar{\rho_S}$ is the
mean density of Jupiter if it was a sphere of radius $a_{1bar} = 71492$ km, and $\Delta a$
is the depth of the high atmosphere.

As mentioned above, $\Delta a \simeq 70$ km and we know that $\bar{\rho_S}\simeq 10^3$ kg m$^{-3}$.
Taking the outer density in a range  $(0.01-1)\times \rho_{1bar}$, and
the conditions of Jupiter, $\rho_{1bar} \simeq 0.17$ kg $m^{-3}$, yields:
\begin{gather}
\phi_{neg} \in [5 \times 10^{-9} ; 5 \times 10^{-7}] \times \dfrac{G M}{a_{1bar}}.
\end{gather}
This is clearly a first order correction since the potential on each spheroid, 
given by the CMS method, ranges from $G M/a_{1bar}$ in the outermost
layers to $2G M/a_{1bar}$ inside. 

The neglected potential does not depend on the layer, so the hydrostatic 
assumption (using the gradient of the potential) remains valid. The main source of change comes from
the shapes of the spheroids. We calculated the impact
of this change in Appendix \ref{app:phi_neg} with the use of equation (40)
of H13 and obtained the following result : 

\begin{equation}
\dfrac{\Delta J_{2,0}}{J_{2,0}}  \simeq 287 \times
\left(\dfrac{\rho_{ext}}{\bar{\rho_S}} \right)
\left(\dfrac{\Delta a}{a_{1bar}}  \right).
\end{equation}
Within the above range of outer densities we get: 
\begin{equation}
\dfrac{\Delta J_{2,0}}{J_{2,0}}  \in [5 \times 10^{-7};5 \times 10^{-5}].
\label{eq:delta_J2,0}
\end{equation}
Since the potential is smallest in the outside layers, we expect the relative change in $J_{2,i}$ of the $i^{th}$ spheroid to decrease with depth. Nevertheless,
we have seen that the CMS method requires a large number of spheroids in the high
atmosphere (a point which is validated in the next section). We thus expect the change in the 
$J_{2,i}$  in the outer layers, which have the largest impact on $J_2$, to be comparable to
Eq.(\ref{eq:delta_J2,0}). 
If we make the bold assumption that every layer
contributes equally and choose for the outer density $\rho_{ext} = 0.03$ kg m$^{-3}$, which
is the value that corresponds to mass conservation in the high atmosphere, we get : 
\begin{equation}
\Delta J_{2}^{ext}\times 10^6 \simeq 8 \times 10^{-6} (J_{2}^{ext} \times 10^6) \simeq 1 \times 10^{-1}.
\label{eq:delta_J2_highatm}
\end{equation}
This is approximately 1/2 Juno error bars. Even though this relies on several assumptions, 
mainly that we can approximate the outer layers by a constant density spheroid
and that the error is the same on each spheroid, it gives a reasonable estimate of the impact of neglecting the high atmosphere
on $J_2$. 
From the evaluation of $J_4$ and $J_6$ in Appendix \ref{app:phi_neg}, we 
calculated that:
\begin{gather}
\Delta J_{4}^{ext}\times 10^6 \simeq 6 \times 10^{-2}, \\
\Delta J_{6}^{ext}\times 10^6 \simeq 2 \times 10^{-2}. 
\label{eq:delta_Jn_highatm}
\end{gather}
We see that the agreement with Juno's error bars improves with 
higher order moments, although it stabilizes around $10^{-2}$ from $J_6$ onward. 
In order to derive physical information from 
the high order moments, however, one needs to have confidence in the evaluation of 
the first moments. Moreover, these estimations rely on many assumptions 
and give only orders of magnitude.

\begin{figure}[ht!]
% \captionsetup{justification=centering}
\begin{subfigure}{.5\textwidth}
  \centering
  \includegraphics[width=1\linewidth]{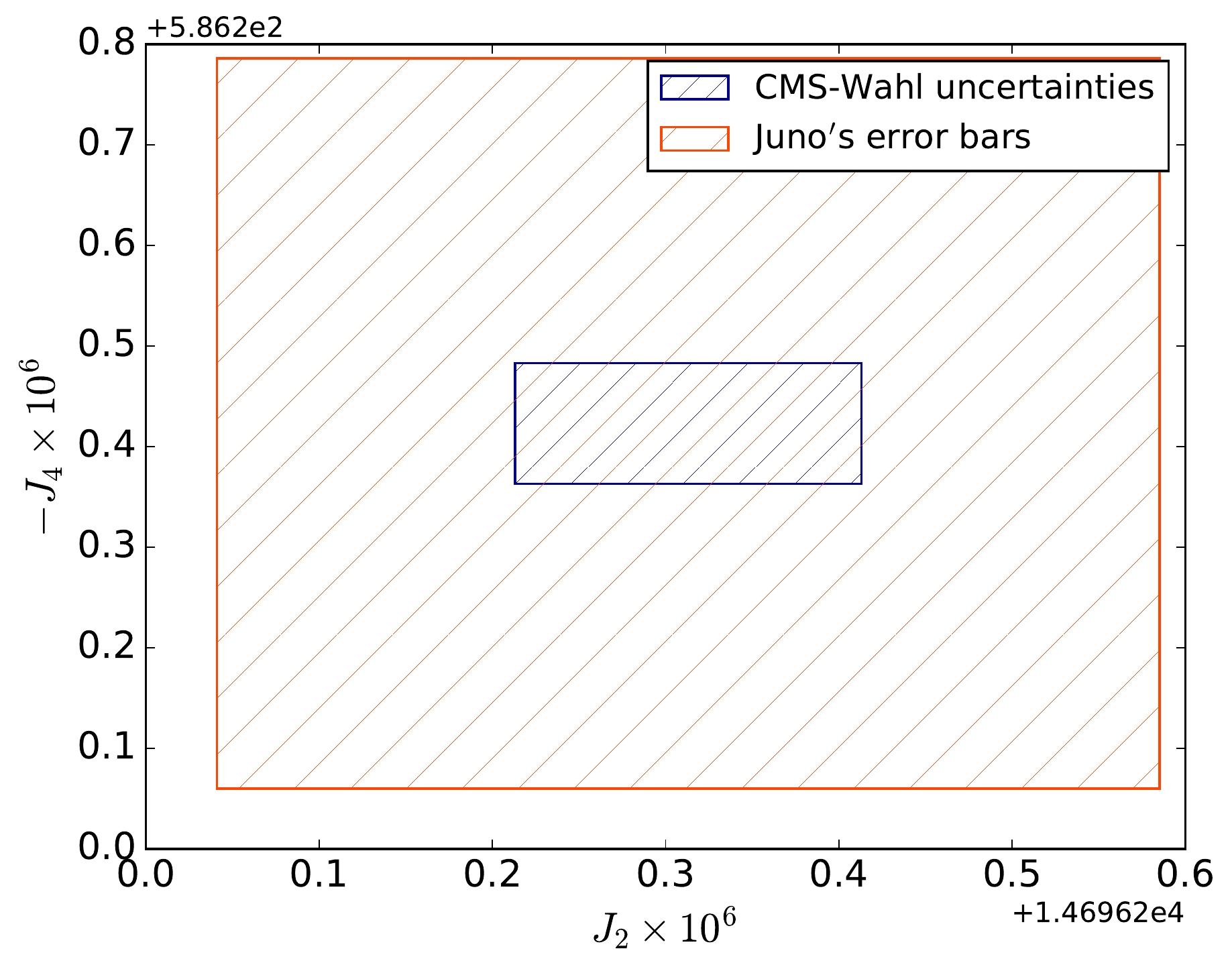}
  \caption{}
\end{subfigure} \\
% \begin{subfigure}{.3\textwidth}
%   \centering
%   \includegraphics[width=1\linewidth]{delta_i_log.pdf}
%   \caption{}
% \end{subfigure} 
\centering 
\begin{subfigure}{.5\textwidth}
  \centering
  \includegraphics[width=1\linewidth]{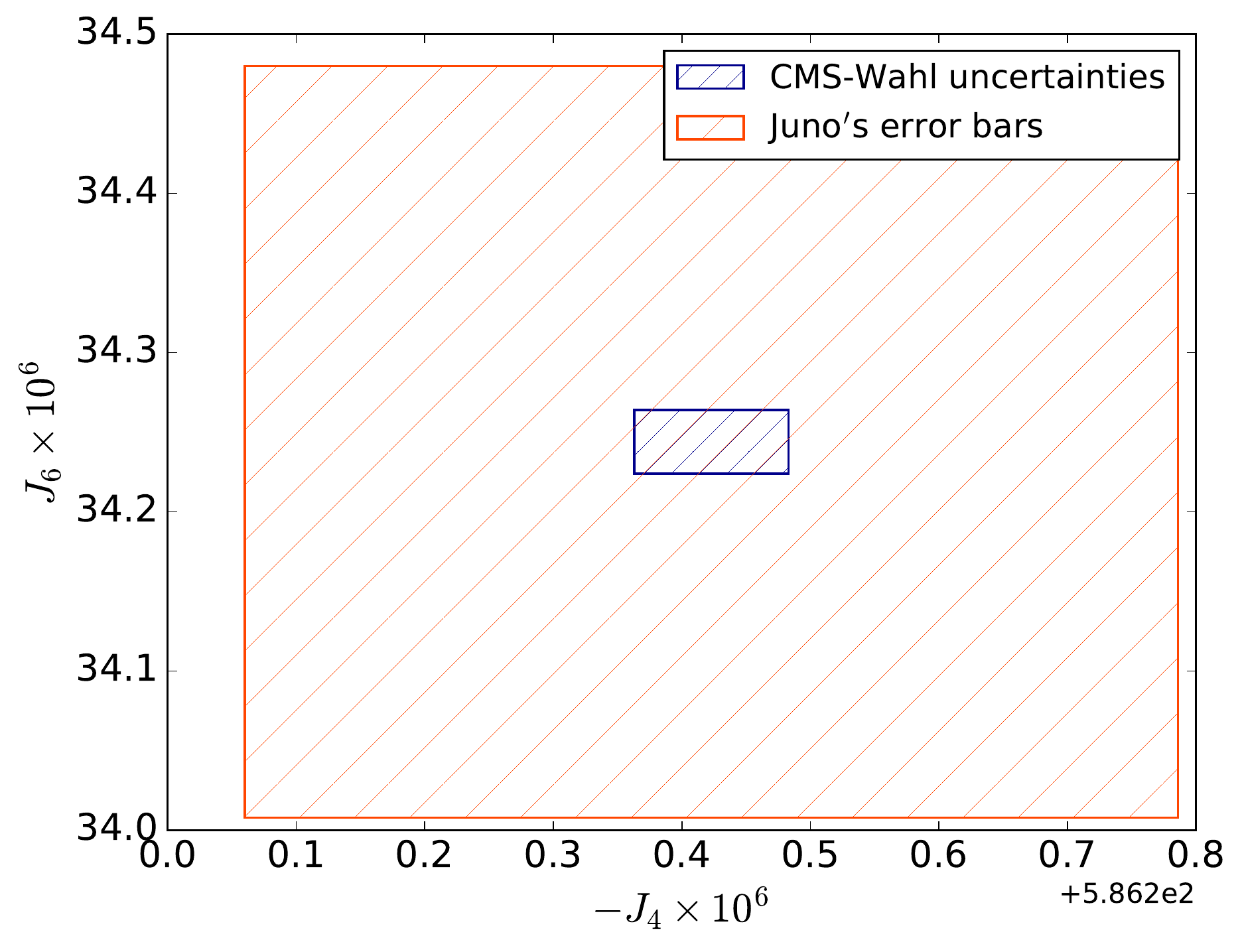}
  \caption{}
  \label{fig:J4J6_wahl}
\end{subfigure}
\caption{Comparison of the errors of the CMS arising 
from neglecting the high atmosphere with Juno's error bar, 
shifted to be centered on the observed $J_n$.
(a) $-J_4 \times 10^6$ vs $J_2 \times 10^6$;
% (b) Same in log scale
(b) $J_6 \times 10^6$ vs $-J_4 \times 10^6$. }
\label{fig:Jn_wahl}
\end{figure}

As in Figure \ref{fig:Jn_Scvh}, we show graphically these uncertainties 
in Figure \ref{fig:Jn_wahl}. We see that we are now able to 
reach Juno's precision. As mentioned in the abstract, however, the analysis 
of future, more accurate data will remain limited by these
intrinsic uncertainties.

If we were able to perform rigorously the calculations of Appendix \ref{app:phi_neg}, 
we would probably find that the errors on the $J_n$'s are correlated, decreasing
the aforederived uncertainties. Unfortunately, a rigorous derivation of these errors would require to relax
many assumptions and would imply very complicated theoretical and numerical calculations. Until this is not done, 
the determination of the $J_n$
are inevitably blurred by the neglected potential, up to half Juno's error bars, as shown
in Figure \ref{fig:Jn_wahl}.

In conclusion, we have shown in this section that the neglected potential due to the high ($<$1 bar) atmosphere region
leads to an error of about half Juno error bars. 
Globally, neglecting the atmosphere above 1 bar in the CMS method leads to
an irreducible error on $J_2\times 10^6$ of the order of $10^{-1}$.
Although such an error level is smaller than Juno's present precision, this will become an issue
if the precision improves further.

It must be stressed that this irreducible error is not 
arising from uncertainties in any physical quantity, such as for example the radius determination
or the eos. It stems from neglecting the high atmospheric levels in the CMS method itself, and
in that regard cannot be removed; indeed, as we have shown in the previous
sections, including the high atmosphere leads to huge uncertainties. Said differently, even with a perfect knowledge of Jupiter radius, internal eos, etc.,
the error, intrinsically linked to the method, remains.
Semi analytical integrations of Eq.(\ref{eq:delta_atmo}) shows
that, globally, this error is of the order of 
$10^{-2}$ for higher order moments. Therefore, the $J_k \times 10^6$ cannot be 
determined within better than a few $10^{-2}$. 
On $J_{10}$ and 
higher order moments, in particular, this error can be close to the order of 
magnitude of the moments themselves. One must
keep that in mind when calculating values of these moments with the CMS method, as in \cite{Wahl17}.

\subsection{Error from the finite number of spheroids}

\begin{figure}[ht!]
\includegraphics[width=\linewidth]{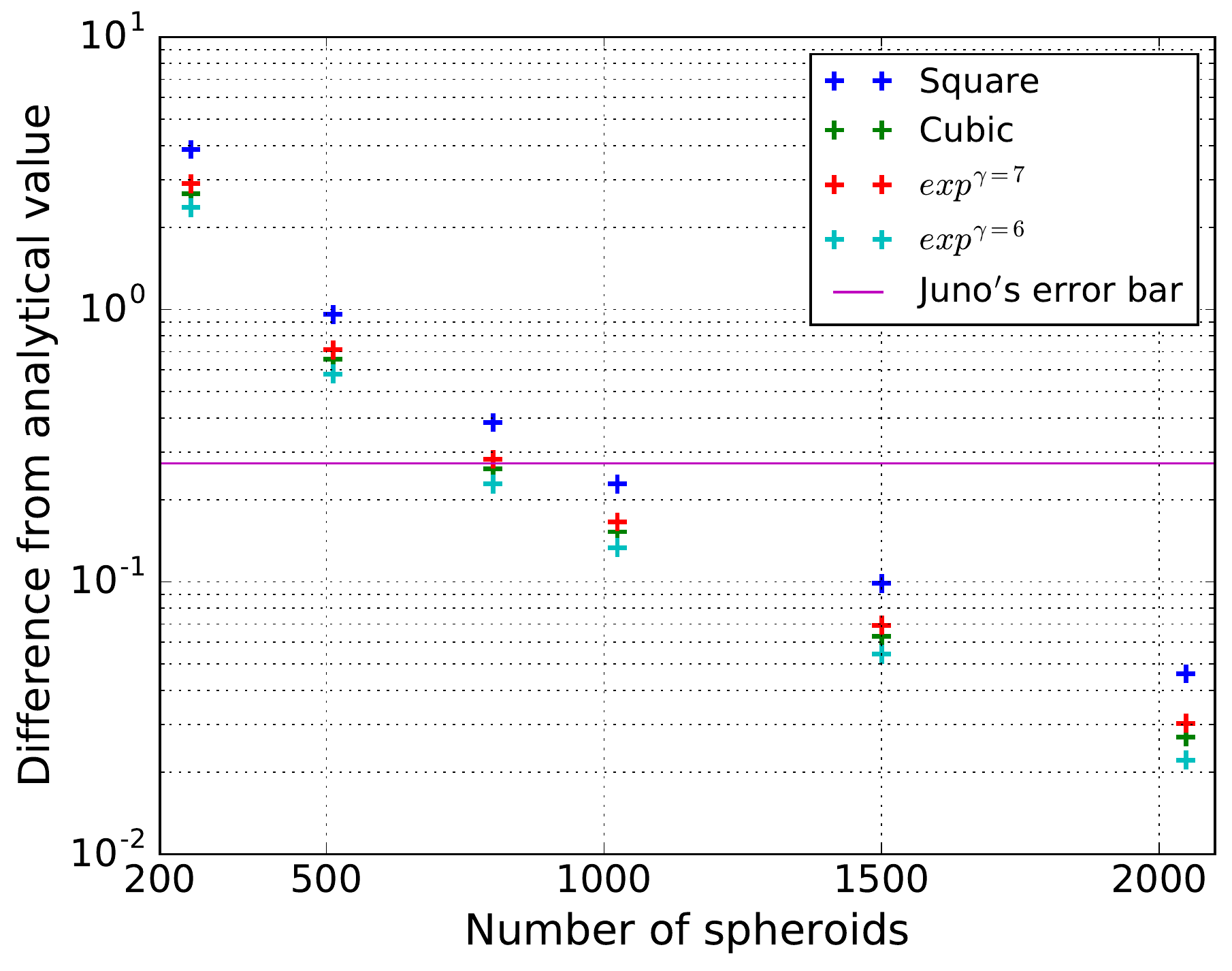}
\caption{Numerical error as a function of the number of spheroids when neglecting the high atmosphere for
a polytropic eos.}
\label{fig:err_highatm}
\end{figure}

Now that we have estimated every possible source of error and showed that 
the method used by \cite{Wahl17} is applicable with the current error bars of the Juno mission,
we need to examine how the uncertainties depend on the number of spheroids and
verify whether the conclusions of \S\ref{numerics} are affected or not
by the change of outer boundary condition. Figure \ref{fig:err_highatm} displays the dependence of the error upon the number
of spheroids for different spheroid repartitions in the present case.

Interestingly, the exponential $\gamma=6$ repartition happens to provide the best 
match to the analytical results. More importantly, we show that
some of the different repartitions give the same results, a mandatory condition to use the method with confidence
with a realistic eos, as mentioned earlier.

We see, again, that 512 spheroids are not enough to satisfy Juno's precision.
As mentioned previously, even though the CMS values are precise at $10^{-13}$, the
discretization error dominates.
Depending on the spheroid repartition, this error can become quite significant. Unfortunately, \cite{Wahl17} do not mention which spheroid repartition they use exactly.
As shown in the Figure, however, this needs to be clearly specified when deriving Jupiter models in the context of Juno to verify their validity.

When neglecting the high atmosphere, the calculations are not hampered by any degeneracy due to the first layers since the densities, 
which are larger than $\rho_{1bar} \simeq 0.17$ kg m$^{-3}$ are well described by the eos.
Therefore, the only errors on the derived quantities of interest come from the discretization errors.
As in \S\ref{ssec : errors_Y}, we give in Table \ref{tab:Y_highatm} the values
of $Y'$, $Y_2'$ and $M_c$ obtained when matching $J_2$ and $J_4$ in the same conditions.
% The first thing to notice is that these values are quite different than from the ones in
One might notice that these results are different from the ones in
Table \ref{Test_Y_mc}; this is only due to the fact that in Table \ref{Test_Y_mc} 
the outer radius is $71492$ km, instead of the radius at 1 bar. We 
checked that if the external radii are consistent, the derived physical quantities
are within the range of uncertainties of \S\ref{sec:SCVH}.

\begin{table}[ht!]
\caption{Same as Table \ref{Test_Y_mc} when neglecting the high atmosphere.
}
\label{tab:Y_highatm}
\begin{center}
\begin{tabular}{c | c || c  c  }
\hline
\hline
N & & Quadratic & Cubic \\
\hline
\cline{3-4}
 \multirow{3}{*}{512} & Y$^\prime$ &  0.3210
 & 0.3211\\
 \cline{2-4}
  & Y$^\prime_2$ &0.372001 & 0.372015 \\
 \cline{2-4}
  & $M_c$  & 9.6849 & 9.6709 \\
  \hline
  \hline
 \multirow{3}{*}{1500} & Y$^\prime$ & 0.3218 & 0.3218 \\
 \cline{2-4}
  & Y$^\prime_2$ & 0.372098 & 0.372090\\
 \cline{2-4}
  & $M_c$ & 9.6546& 9.6624 \\
\hline
\hline
\end{tabular}
\end{center}
\end{table}

The differences from 512 to 1500 spheroids are of the order of $0.3\%$ on the physical quantities.
Because of the degeneracy of models leading to the same external gravitational moments (by e.g., changing the
pressure of the entropy jump, the amount of metals, etc..) we do not expect to derive
a unique model with such a precision on the core mass, so, in this context, 512 spheroids can be considered as sufficient
to derive accurate enough models of Jupiter. When using the physical quantities obtained with
512 spheroids in calculations done with 1500 spheroids, we get an error on $J_2 \times 10^6$ of $\sim 1$, 
$10\times$ Juno error bars. Therefore, even though one can derive a precise enough model with 512 spheroids, it is highly recommended, as a sanity check,
to verify with 1500 whether Juno precision can be reached with only a slight change of the derived physical quantities.

We still have to evaluate the errors coming from neglecting the high atmosphere. To evaluate
the largest possible uncertainty, we added $1 \times 10^{-1}$ to $J_2 \times 10^6$ and removed it from $J_4\times 10^6$.
We stress again that this must be done
because the value of both $J_2$ and $J_4$ can not be trusted under the aforederived intrinsic irreducible error
of the method, due to the omission of the high atmosphere, even when fitting to observed values at much higher precision. The results are given in Table
\ref{tab:Y_highatm_+1-1}.
The differences between Table \ref{tab:Y_highatm} and \ref{tab:Y_highatm_+1-1} show that
neglecting the high atmosphere yields an uncertainty on the physical
quantities of a few parts in a thousand, even up to 2 to 3\% on the core mass. 

\begin{table}[ht!]
\caption{Same as Table \ref{tab:Y_highatm} except that $J_2 \times 10^6$
and $J_4 \times 10^6$ have been changed by $ \pm 1 \times 10^{-1}$. +1-1 means that we add
$1 \times 10^{-1}$ on $J_2$ and remove it from $J_4$, -1+1 is the opposite.
}
\label{tab:Y_highatm_+1-1}
\begin{center}
\begin{tabular}{c | c || c  c  }
\hline
\hline
N & & Cubic +1-1 & Cubic -1+1 \\
\hline
\cline{3-4}
 \multirow{3}{*}{512} & Y$^\prime$ &  0.3212
 & 0.3225\\
 \cline{2-4}
  & Y$^\prime_2$ &0.372155 &0.372024 \\
 \cline{2-4}
  & $M_c$  & 9.5297 & 9.8015 \\
  \hline
  \hline
 \multirow{3}{*}{1500} & Y$^\prime$ & 0.3212 &  0.3225 \\
 \cline{2-4}
  & Y$^\prime_2$ & 0.372149 & 0.372022\\
 \cline{2-4}
  & $M_c$ & 9.5374 & 9.8053\\
\hline
\hline
\end{tabular}
\end{center}
\end{table}

%We here studied the worst cases and the way we deal with the core is very simplified (constant density), we can thus expect a smaller uncertainty of around one percent on the core and a few parts per thousand on the Helium mass fractions.

This is the irreducible error of the CMS method when neglecting the levels of the high atmosphere ($<$1 bar). An improvement
in the precision of Juno cannot change this uncertainty. In that regard, 
using more than 1500 spheroids would be a second order correction on the 
method uncertainty.

\section{Conclusion}
\label{conclusion}

In this paper, we have examined under which
configuration the CMS method can be safely used in the context of the Juno mission.
 We have first derived analytical expressions, based on a polytropic eos,
to evaluate the errors in the calculation of the gravitational moments with the CMS method in various cases.
These analytical derivations relied on simplifying approximations and 
thus represent lower limits of the errors. When 
applied to calculations based on 512 spheroids, as done for instance in the works of \cite{HM2016}
these expressions show
that it is not possible to match the Juno's constraints because of the discretization error. 
To verify the analytical calculations, we have developed a code based on the method exposed
by \cite{Hub2013}. The numerical calculations confirmed the analytical results
and also showed that the external
layers of the planet have a huge impact of the determination of $J_2$ and need to be described with very high accuracy. 
At first, this appeared to be the biggest limitation of the CMS method: 
for a precise evaluation of the gravitational moments 
of Jupiter, this method requires layers in the high
atmosphere smaller than 1 km. However, this implies densities so low that the very concept of an equilibrium
equation of state to describe these layers becomes questionable, while density profiles fit from observations would not solve
this issue (as explained at the end of \S \ref{first_layers}).

We have shown, with a polytrope $n=1$, that one needs a quadratic or exponential 
repartition of at least 1500 spheroids to reach precise enough values of the moments to match Juno's constraints, which implies
densities of the order 
of a few 10$^{-2}$ kg m$^{-3}$ for the outermost layers. With such specifications, the CMS method 
can in principle reach precisions better than Juno's uncertainties.

When these calculations are performed numerically with barotropes prescribed from an appropriate eos for the external layers, 
the above conclusions were confirmed: the CMS method requires
a square, cubic or exponential spacing of at least 1500 spheroids. None of these calculations, 
however, was able to get a satisfying description
of the first layers. Indeed, even though each of these individual layers contributes negligibly to $J_2$,
a small change in the density structure of these regions
leads eventually to unacceptable changes in the total value of $J_2$. 
Furthermore, when fitting $J_2$ and $J_4$, we have shown that the description of these layers 
leads to a degeneracy of solutions which hampers the determination of physical quantities such as the
heavy element mass fraction or the mass of the core to better than a few percents. 

In order to overcome this issue, \cite{Wahl17} propose to neglect the 
high atmosphere and impose the 1 bar pressure as an outer boundary condition. We have evaluated
the errors coming from this assumption, and showed that they are of the order of half Juno error bars. 
In principle, the CMS method with this boundary condition can thus be applied to 
derive Jupiter models within the current precision of Juno, with an irreducible
uncertainty arising from the neglected high atmosphere. This uncertainty is of the order 
of 0.1 on $J_2 \times 10^{6}$ while the current precision of Juno is $ \pm 0.272$. The
remaining sources of errors
on the physical quantities, however, stem from discretization errors, and 512 spheroids can give results with a 
precision of a few parts of a thousand. Considering the degeneracy of models that could fit 
with Jupiter's observed gravitational moments, this can be considered as an acceptable uncertainty. However, we show that,
due to the aforementioned irreducible error of the method, increasing Juno precision will not enable us to
derive more precise models. Quantitatively, the irreducible errors yield global uncertainties on
the physical quantities derived in a model close to 1\%.

The conclusion of this study is that the Concentric MacLaurin Spheroid method such as used in HM16 could not satisfy Juno's constraints and needed to be improved
to derive accurate enough Jupiter model, invalidating any model up to this stage in the context of the Juno mission. First, a larger number of spheroids with a better
repartition is mandatory. Second, there is no satisfying solution to safely take into account the impact of the high atmosphere region above 1 bar.
We have shown that imposing the 1 bar radius as the outer boundary condition, as done in \cite{Wahl17}, is acceptable,
although it leads to irreducible errors of a few parts of a thousand on the derived physical quantities, increased
by an eventual discretization error. Although such a precision can be considered as satisfactory, given all the other sources of uncertainty in the input physics of the model, 
this shows that, if the precision on Juno's data improves, the CMS method in
its present form will not be able to exploit this improvement to refine further the models.
\\

{\it Acknowledgements}
% \begin{acknowledgements}
   The authors are deeply indebted to W. Hubbard for kindly providing a version of his CMS code, which enabled us to carefully compare our results with the published ones, and for useful conversations. 
% \end{acknowledgements}

\bibliography{biblio_DC17_arxiv}

\appendix
\section{Simplifications for the angular part}
\label{mu_legendre}

Here, we show in details how we got rid of the angular 
part in Eq.(\ref{eq:def_deltaJ2k}). The integral we
wanted to simplify is: 

\begin{equation}
II = \int_{0}^{1}\int_{0}^{a_J}
\rho (r') r'^{2k+2} P_{2k}(\mu) dr' d\mu .
\end{equation}
We approached this integral differently: instead of just
considering an integral over $r'$ and $\mu$, we 
rather followed the surfaces of constant density. Then, $r'(\mu)$ was
defined in order to have $\rho(r'(\mu)) = cst$ when $\mu$ varies
from 0 to 1. Another way to say it is that we took the value of $\rho$ 
for a given equatorial radius $r_{eq}$, and calculated the integral over constant 
$\rho$ from equator to pole. 

In order to go further, we considered that the surfaces of constant potential
were ellipsoids with the same ratio of polar to equatorial radius $e$. That gave, 
for every equipotential surface: 
\begin{equation}
r'(\mu) = \dfrac{r_{eq}}{\sqrt{1+e^2\mu^2}},
\label{r_of_mu}
\end{equation}
where $r_{eq}$ is the equatorial radius of the spheroid.
In reality, this is not exactly the case, the isobaric shperoids are not
ellipsoids and the innermost ones are less flattened than the outermost ones, but it gives
an idea of the integral over $\mu$.

We just had to change variables : $\mu \rightarrow \mu$ and $r' \rightarrow r_{eq}/\sqrt{1+e^2\mu^2}$.
With $\mu$ varying from $0$ to $1$ and $r_{eq}$ from $0$ to $a_J$, we have not changed the 
domain of integration because outside of the outermost ellipsoid - in our ellipsoidal approximation - 
we had $\rho (a_J, \mu >0) = 0$.
The Jacobian is $1/\sqrt{1+e^2 \mu^2}$ and we obtained : 
\begin{equation}
II = \int_{0}^{a_J} \rho (r_{eq})r_{eq}^{2k+2}  \int_{0}^{1}
\dfrac{P_{2k}(\mu)}{\left(\sqrt{1+e^2\mu^2}\right)^{2k+3}}  d\mu\, dr_{eq}.
\end{equation}

From Eq.(\ref{density_ZT}), 
we see that the density only depends on $l/l_0$, the mean radius 
of one isobaric layer above the external mean radius.
With Eq.(\ref{r_of_mu}), we were then able to write $l = \beta \times r_{eq}$
with the same $\beta$ for all $r_{eq}$. The $\beta$ cancelled in the fraction, 
and we simply changed $l/l_0$ by 
$r_{eq}/a_J$. \\

We just had to calculate 
the other part of the integral, with the analytical expression of Eq.(\ref{r_of_mu}) :
\begin{equation}
<P_{2k}> =  \int_{0}^{1}
\left(\dfrac{1}{1+e^2\mu^2}\right)^{\frac{2k+3}{2}} P_{2k}(\mu)  d\mu.
\label{def_<P>}
\end{equation}
Since there is no simple way to do it, we kept it
in our equations, 
but the interesting aspect is that it no longer depends on $r'$: 
we have reduced our equation to a simple integral.

We were able to check whether this is correct or not. So far,
we have supposed that the equipotential surfaces are ellipsoids
and that they have the same $e$ value. Considering 
that we integrate along these surfaces by varying the equatorial 
radius from the center to $a_J$, we showed that we can write:

\begin{eqnarray}
J_{2k} \sim - \dfrac{4\pi}{Ma_J^{2k}}
\left( \int_{0}^{1}
\left(\dfrac{1}{1+e^2\mu^2}\right)^{\frac{2k+3}{2}} P_{2k}(\mu)  d\mu \right) 
\nonumber \\
\times \int_0^{a_J} \rho(r_{eq}) r_{eq}^{2k+2} dr_{eq}. \text{ }
\label{eval_J2}
\end{eqnarray}
It was easy to calculate these integrals as we know
their analytical expression. We just needed to choose the correct $e$ and $m$ values. 
From the observed equatorial and polar radii of Jupiter (see Table \ref{values_Jupiter}),
we have :
\begin{equation}
e = \sqrt{71492^2/66854^2-1} \approx 0.378897
\end{equation}
These radii also give us the value for the mean density : 
$\bar{\rho} = \dfrac{M_J}{\dfrac{4}{3}\pi a_J^2 (a_J)_{polaire}}
\approx 1326.5$ kg m$^{-3}$, which allowed us to calculate 
\begin{equation}
m = \dfrac{3 \omega^2}{4 \pi G \bar{\rho}} \approx 0.083408
\end{equation}
Implemented in Eq.(\ref{eval_J2}) with the 
expression of density from Eq.(\ref{density_ZT}), a numerical 
integration gave: 
\begin{gather}
J_2 \times 10^6 = 16348.804 \text{  with  } 
J_2^{theory} \times 10^6 = 13988.511  \nonumber \\
-J_4\times 10^6   = 643.983 \text{  with  } 
-J_4^{theory}\times 10^6  = 531.828  \nonumber \\
J_6\times 10^6  = 35.16 \text{  with  } 
J_6^{theory}\times 10^6  = 30.12
\end{gather}

There is about $20 \%$ difference with the theoretical results, which is expected 
as we have chosen the outermost $e$ value, which is the largest.
With $e = 0.36$, we obtained less than $10 \%$ difference on these moments.
Considering further approximations that are made in the next appendices, 
a $0.9$ factor is acceptable to compare the theory with the numerical
results and we retain this $e = 0.36$ value in our calculations.

\section{Supplementary calculations for the general case}
\label{maths}

Going back to Eq.(\ref{eq:def_deltaJ2k}), Eq.(\ref{density_ZT}) and Eq.(\ref{def_<P>}) we have : 
\begin{gather}
\left| \Delta J_{2k} \right| \sim \bigg| \dfrac{4\pi}{Ma_J^{2k}}\sum_{i=0}^{N-1} <P_{2k}> 
\nonumber \\
\times \int_{r_{i+1}}^{r_i}
\bar{A}\left[\dfrac{sin(\alpha \dfrac{r'}{a_J})}{\dfrac{r'}{a_J}}
- \dfrac{sin(\alpha \dfrac{R_i}{a_J})}{\dfrac{R_i}{a_J}}
\right]
r'^{2k+2} dr' \label{J2k_maths} \bigg|,\\
{\rm with}\,\,R_i = \dfrac{r_{i+1}+r_{i}}{2}, \label{def_RA}\\
{\rm and}\,\,\bar{A} = \bar{\rho} A \label{def_Abar}.
\end{gather}

Because there is a high number of layers, we were able to approximate
$|R_i-r'| \ll R_i$ everywhere. This approximation
is less valid in the $\sim 10$ deepest layers, but 
their impact on the gravitational moments is 
negligible (Figure \ref{impact_J2}). 

With : 
\begin{equation}
\gamma = \dfrac{\alpha}{a_J},
\end{equation}
we developped the sinus to order 2 : 
\begin{gather}
%\begin{eqnarray}
\dfrac{\sin\left( \gamma r' \right)}{r'} = 
\dfrac{\sin\left( \gamma R_i\left(1 + \dfrac{r'-R_i}{R_i} \right) \right)}
{R_i \left(1+\dfrac{r'-R_i}{R_i} \right)} \nonumber \\
\dfrac{\sin\left( \gamma r' \right)}{r'}  \simeq \dfrac{1}{R_i}
\left(1-\dfrac{\xi}{R_i} +\left(\dfrac{\xi}{R_i}\right)^2 \right) 
\bigg(\sin(\gamma R_i)\cos(\gamma \xi) + \nonumber \\ 
\,\,\,\,\,\,\,\,\,\,\,\,\,\,\,\,\,\, \cos(\gamma R_i) \sin(\gamma \xi) \bigg),\nonumber \\
\text{      where      } 
\xi = r'-R_i \ll R_i. \nonumber
\end{gather}
%\end{eqnarray}
We know that $\alpha \sim \pi$. In the internal layers, $a_J \gg R_i$,
so $\gamma \xi =  \xi \alpha / a_J \ll \xi /R_i$. 
On the outside, $a_J \sim R_i $ so 
$\gamma \xi  \sim  \xi /R_i $. As mentioned above,  $ \xi / R_i \ll 1$,
which means that everywhere  $\gamma \xi \ll 1$. Therefore: 

\begin{gather}
\dfrac{\sin\left( \gamma r' \right)}{r'} \simeq \dfrac{1}{R_i}
\left(1-\dfrac{\xi}{R_i} +\left(\dfrac{\xi}{R_i}\right)^2 \right) 
\bigg(\sin(\gamma R_i)\left(1 - \dfrac{(\gamma \xi)^2}{2} \right) \nonumber \\
+ \cos(\gamma R_i)
\left( \gamma\xi \right)\bigg), \nonumber \\
\dfrac{\sin\left( \gamma r' \right)}{r'} \simeq \dfrac{\sin(\gamma R_i)}{R_i} + 
\xi \left( \dfrac{\gamma}{R_i} \cos(\gamma R_i) - \dfrac{1}{R_i^2}\sin(\gamma R_i) \right) +
\nonumber \\
\xi^2 \left( - \dfrac{\gamma^2}{2R_i}\sin(\gamma R_i) -
\dfrac{\gamma}{R_i^2} \cos(\gamma R_i) + \dfrac{1}{R_i^3} \sin(\gamma R_i)   \right).
\end{gather}
The zeroth order term cancelled in Eq.(\ref{J2k_maths})
so :

\begin{gather}
|\Delta J_{2k}| \sim \bigg| \dfrac{4\pi}{Ma_J^{2k}} \displaystyle{\sum_{i=0}^{N-1}}
 <P_{2k}>\dfrac{\bar{A}a_J}{R_i} \nonumber \\
\times \displaystyle{\int_{r_{i+1}}^{r_i}}
\left[ (r' - R_i) \left( \gamma \cos(\gamma R_i) - \dfrac{1}{R_i}\sin(\gamma R_i) \right) + 
\right. \nonumber \\ \left.
(r' - R_i)^2 \left( - \dfrac{\gamma^2}{2}\sin(\gamma R_i) -
\dfrac{\gamma}{R_i} \cos(\gamma R_i) + \dfrac{1}{R_i^2} \sin(\gamma R_i)   \right) \right. \nonumber \\ \left.
\right]r'^{2k+2} dr'\bigg|. 
\label{potential}
\end{gather}

We introduced
\begin{gather}
C_1 = \left( \gamma \cos(\gamma R_i) - \dfrac{1}{R_i}\sin(\gamma R_i) \right) \nonumber \\
C_2 = \left( - \dfrac{\gamma^2}{2}\sin(\gamma R_i) -
\dfrac{\gamma}{R_i} \cos(\gamma R_i) + \dfrac{1}{R_i^2} \sin(\gamma R_i)   \right). \nonumber \\
\end{gather}
This gave a simplified expression for the potential : 

\begin{gather}
|\Delta J_{2k}| \sim \bigg| \dfrac{4\pi}{Ma_J^{2k}} 
\sum_{i=0}^{N-1} <P_{2k}>\dfrac{\bar{A}a_J}{R_i} \nonumber \\
\times \displaystyle{\int_{r_{i+1}}^{r_i}}
\left (C_1 (r' - R_i) + C_2
(r' - R_i)^2  \right)r'^{2k+2} dr' \bigg|
\end{gather}
Developping the power of $r'$ yielded:

\begin{gather}
r'^{2k+2} = R_i^{2k+2} \left(1+\dfrac{r'-R_i}{R_i} \right)^{2k+2}, \nonumber \\
r'^{2k+2} \simeq 
R_i^{2k+2} \left(1 + (2k+2)\dfrac{r'-R_i}{R_i} \right).
\end{gather}
Then, the integral $I$ is given by :
\begin{gather}
I \simeq R_i^{2k+2} \displaystyle{\int_{r_{i+1}}^{r_i}}
\bigg[C_1 (r' - R_i) + \nonumber \\ 
\,\,\,\,\,\,\,\,\,\,\,\,\,\,\,\,\,\, \left( C_2 + (2k+2) \dfrac{C_1}{R_i} \right)
(r' - R_i)^2 \bigg] dr'.
\end{gather}
We changed the variable of integration $r' \rightarrow (r'-R_i)$,
and using Eq.(\ref{def_RA}) : 

\begin{equation}
I \simeq R_i^{2k+2} \displaystyle{\int_{\frac{r_{i+1}-r_i}{2}}^{\frac{r_i-r_{i+1}}{2}}}
\left [C_1 r' + \left( C_2 + (2k+2) \dfrac{C_1}{R_i} \right)
r'^2 \right] dr'.
\end{equation}
As a first calculation, we chose a constant $\Delta r$ with
depth, as suggested in H13,
$\Delta r =a_J /N$. If one wants another repartition
of spheroids, they just have to change $\Delta r$ by $\Delta r_i$ depending on
depth: 

\begin{equation}
I \simeq R_i^{2k+2} \displaystyle{\int_{-\frac{\Delta r}{2}}^{\frac{\Delta r}{2}}}
\left [C_1 r' + \left( C_2 + (2k+2) \dfrac{C_1}{R_i} \right)
r'^2 \right] dr'.
\end{equation}
This is easily calculated and we got:

\begin{gather}
I \simeq R_i^{2k+2} \, \left( C_2 + (2k+2) \dfrac{C_1}{R_i} \right)
\dfrac{\Delta r^3}{12}.
\end{gather}
Puting this expression into Eq.(\ref{potential}), we obtained an intermediate formula : 

\begin{align}
|\Delta J_{2k}| \sim \bigg| \dfrac{4\pi}{Ma_J^{2k}}
\sum_{i=0}^{N-1} <P_{2k}>\bar{A}a_J R_i^{2k+1} \nonumber \\
\left( C_2 + (2k+2) \dfrac{C_1}{R_i} \right)
\dfrac{\Delta r^3}{12} \bigg|.
\label{moments_maths}
\end{align}

\begin{figure}[ht!]
% \captionsetup{justification=centering}
\begin{subfigure}{.5\textwidth}
  \centering
  \includegraphics[width=1\linewidth, height=8cm]{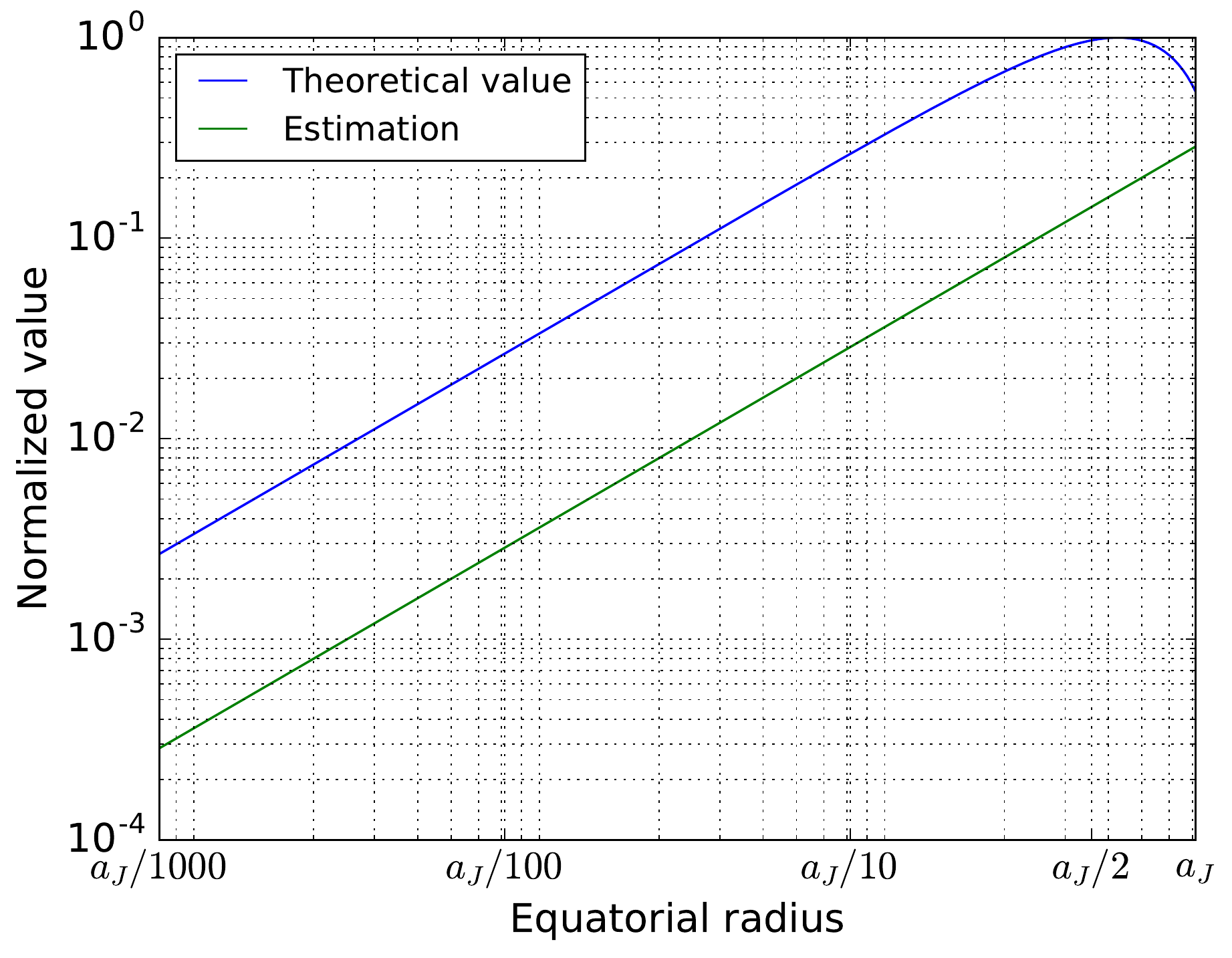}
\end{subfigure}% 
\\
\begin{subfigure}{.5\textwidth}
  \flushright
  \includegraphics[width=1\linewidth, height=8cm]{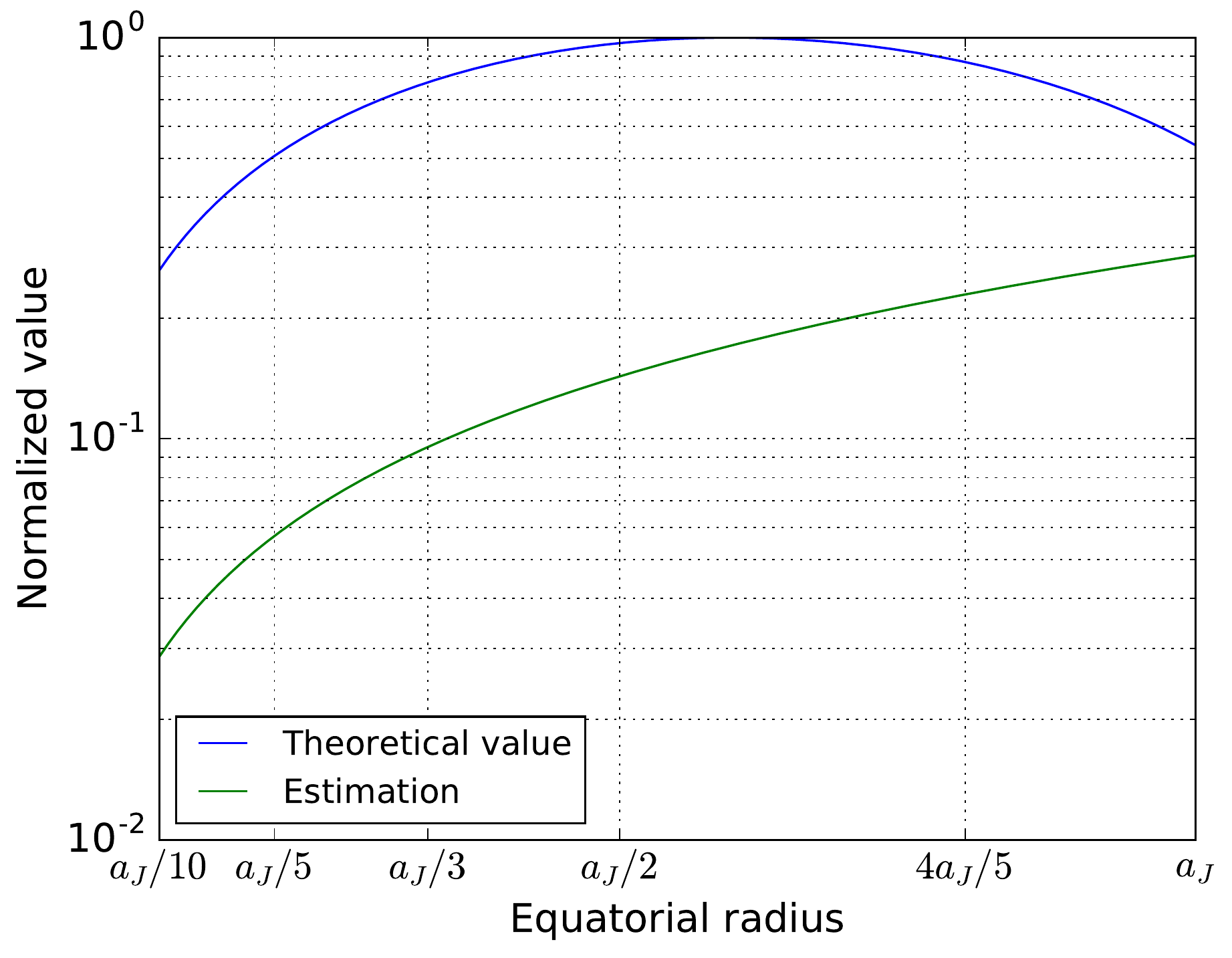}
\end{subfigure}
\caption{Normalized value of $C2+4C1/R_i$ and 
of its approximation $5R_i/a_J^3$ (Eq.(\ref{C1_C2})) as a function of 
altitude $R_i$. 
(a) From center to exterior;
(b) zoom on the external layers }
\label{check_C1C2}
\end{figure}

\noindent We considered two cases: in the internal layers, where $a_J \gg R_i$: 

\begin{gather}
C_1 \sim \gamma -\dfrac{\gamma ^3 R_i^2}{2} - \gamma + \dfrac{\gamma^3 R_i^2}{6} 
= -\dfrac{\alpha^3}{3} \dfrac{R_i^2}{a_j^3}, \\
C_2 \sim -\dfrac{\gamma^3 R_i}{2} - \dfrac{\gamma}{R_i} + \dfrac{\gamma^3 R_i}{2}
+ \dfrac{\gamma}{R_i} - \dfrac{\gamma^3 R_i}{6} = - \dfrac{\alpha^3}{6}\dfrac{R_i}{a_j^3}.
\end{gather}
In the external layers, $a_J \approx R_i$, there is no obvious approximation. We assumed
that $C_1$ and $C_2$ were of the same order of magnitude as the coefficient in front
of the sinusoidal functions and, remembering that $R_i/a_J \sim 1$:
\begin{gather}
C_1 \sim \gamma \sim \dfrac{1}{R_i} = \left(\dfrac{R_i}{a_j}\right)^2 \dfrac{1}{a_J} =  \dfrac{R_i^2}{a_j^3}, \\
C_2 \sim \gamma^2 \sim \dfrac{1}{R_i^2} = \dfrac{R_i}{a_J}\dfrac{1}{a_J^2} = \dfrac{R_i}{a_j^3}.
\end{gather}
Neglecting the factor $\alpha^3 /3$ (which varies with height):

\begin{gather}
C_1 \sim \dfrac{R_i^2}{a_j^3}  \nonumber \\
C_2 \sim \dfrac{R_i}{a_j^3},
%\label{C1_C2}
\end{gather}
which yielded the aproximated relation:
\begin{gather}
C_2 + (2k+2) \dfrac{C_1}{R_i} \sim (2k+3)\dfrac{R_i}{a_J^3}.
\label{C1_C2}
\end{gather}
%So far, we have only used approximations that would increase the error
%if they were not considered (perfect shape
%and density for the spheroids). 
In Figure \ref{check_C1C2},
we plot the exact and approximated values of $C2+(2k+2)C1$ for $k=1$.
As expected, in the interior we recover the factor ten shift ($\alpha^3/3$) between
the real and estimated value, whereas
in the externalmost layers this term is smaller so we underestimate the 
error by a factor $\sim 2-5$. Everything is thus consistent
with our assumptions. 

With these assumptions, we were then able to write (redefining $<P_{2k}>$ as its absolute value) : 
\begin{equation}
|\Delta J_{2k}| \sim 4\pi(2k+3)\dfrac{\Delta r^3}{12}\dfrac{\bar{A}<P_{2k}>}{M}
\sum_{i=0}^{N-1} \left( \dfrac{R_i}{a_J}\right)^{2k+2}.
\label{J_2k_deltar}
\end{equation}
It is important to note that up to now, if $\Delta r$ was not constant
we would still have the same result by using $\Delta r_i$ instead of $\Delta r$
in the sum.
This is done in Appendix \ref{maths_cubic}

In the case of a linear spacing of the spheroids, $\Delta r = a_J/N$, we went further: 
\begin{equation}
\dfrac{\Delta r^3\bar{A}}{M} = \dfrac{a_J^3}{N^3}\dfrac{A \bar{\rho}}{M} = 
\left(\dfrac{a_J^3 \bar{\rho}}{M} \right)\dfrac{A}{N^3} =\dfrac{3}{4 \pi}\dfrac{A}{N^3} \sim \dfrac{1}{4N^3},
\label{simp_deltar}
\end{equation}
because, from Eq.(\ref{def_RA}), $A \sim \pi/3$ if $m\ll 1$, which is valid for Jupiter (and for the
vast majority of celestial bodies).

\noindent As $r_{i+1} = r_i - \Delta r$ and $r_0 = a_J$, $r_i = a_J-i\times\Delta_r$, we get:

\begin{equation}
R_i = \dfrac{r_i+r_{i+1}}{2} = a_J - \left(i+\dfrac{1}{2} \right) \Delta r = a_J \left(1 - \left(i+\dfrac{1}{2} \right)/N\right).
\label{simp_Ri}
\end{equation}
Combining Eq.(\ref{J_2k_deltar}), Eq.(\ref{simp_deltar}) and Eq.(\ref{simp_Ri}) gave the final result, as
written in Eq.(\ref{J_2k_final}) : 
\begin{equation}
%\boxed{
|\Delta J_{2k}| \sim (2k+3)\dfrac{\pi}{12}\dfrac{<P_{2k}>}{N^3}
\sum_{i=0}^{N-1} \left(1 - \left(i+\dfrac{1}{2} \right)/N\right)^{2k+2} 
%}
\end{equation}

\section{Supplementary calculations for J2}
\label{maths_J2}
The formula for $J_2$ is directly derived from Eq.(\ref{J_2k_final}) : 

\begin{equation}
|\Delta J_{2}| \sim \dfrac{5\pi}{12}\dfrac{<P_{2}>}{N^3}
\sum_{i=0}^{N-1} \left(1 - \left(i+\dfrac{1}{2} \right)/N\right)^{4}.
\label{J_2}
\end{equation}

\subsection{Linear spacing}
First, we wanted to calculate $\langle P_2\rangle$ from Eq.(\ref{def_<P>}).
A simple numerical integration yielded (in absolute value): 
\begin{gather}
<P_2> \approx 0.035982
\label{P_2_moyen}
\end{gather}
We could have expanded the sum in Eq.(\ref{J_2}) with the binomial theorem, 
but it is easy to show that the first order is
equivalent to neglecting the factor $1/2$ in the term $(i+1/2)$. So we had 
directly:
\begin{equation}
P  = \left(1 - \left(i+\dfrac{1}{2} \right)/N\right)^{4}
\sim 1 - \dfrac{4}{N}i + \dfrac{6}{N^2}i^2 - \dfrac{4}{N^3}i^3 + \dfrac{i^4}{N^4}.
\label{P_dominant}
\end{equation}
As Eq.(\ref{P_dominant}) corresponds to the 1st order expansion in $i$ in Eq.(\ref{J_2}), we needed
to make sure that these terms did not cancel in the sum for this approximation to be correct.
This sum is 

\begin{gather}
\sum_{i=0}^{N-1} P \approx N - \dfrac{N(N-1)}{2} \dfrac{4}{N} + \dfrac{(N-1)N(2N-1)}{6}\dfrac{6}{N^2} - \nonumber \\
\dfrac{(N-1)^2N^2}{4}\dfrac{4}{N^3}i^3 + \nonumber \\
\dfrac{(N-1)(N)(2N-1)(3(N-1)^2+3(N-1)-1)}{30N^4}. \nonumber \\
\sum_{i=0}^{N-1} P \approx N - 2N +2N - N + \dfrac{N}{5}, \nonumber \\
\text{  so  }\,\,
\sum_{i=0}^{N-1}\left(1 - \left(i+\dfrac{1}{2} \right)/N\right)^{4} \sim \dfrac{N}{5}. \label{P_fin}
\end{gather}
If we wanted to keep every term up to this stage, we would find
that the second order term is $1/2 \ll N/5$. Our approximations
were thus justified.
Finally, Eq.(\ref{J_2}), Eq.(\ref{P_2_moyen}) and Eq.(\ref{P_fin}) yielded: 
\begin{equation}
%\boxed{|\Delta J_{2}| \sim \dfrac{\pi}{12} \dfrac{<P_2>}{N^2} \approx \dfrac{8.8\times 10^{-3}}{N^2}}
|\Delta J_{2}| \sim \dfrac{\pi}{12} \dfrac{<P_2>}{N^2} \approx \dfrac{9.42\times 10^{-3}}{N^2}.
%}
\end{equation}

\subsection{Hubbard and Militzer spacing}
\label{maths_hub}

In HM16, the spacing is not exactly linear. The planet 
is separated in two domains: first, from the outside boundary to $r_i=0.5\, a_J$, they 
use 341 spheroids, and 171 inside this limit. Then, outside 
$\Delta r^{ext} = 3\Delta r/4 $ and inside $\Delta r^{in} = 3\Delta r/2$. 
Moreover, the first spheroid
is a bit particular as it has a a zero density over half its size. 
We did not consider it here: since it is just one spheroid, the impact
on the analytical derivation of the error is small. 

\noindent Using Eq.(\ref{J_2k_deltar}), we were able to write it with a general number $N$ as : 
\begin{gather}
|\Delta J_{2k}| \sim \dfrac{4\pi(2k+3)}{12}\dfrac{\bar{A}<P_{2k}>}{M} \nonumber \\
\times \left[\sum_{i=0}^{\frac{2N}{3}} \left(\dfrac{3 \Delta r}{4}\right)^3
\left( \dfrac{R_i^{ext}}{a_J}\right)^{2k+2} + \right.\nonumber \\ \left.
\sum_{i=\frac{2N}{3}+1}^{N-1} \left(\dfrac{3 \Delta r}{2}\right)^3
\left( \dfrac{R_i^{int}}{a_J}\right)^{2k+2}\right]. \\
R_i^{ext} = a_J \left(1- \dfrac{3}{4N}\left(i+\dfrac{1}{2}\right)\right), \\
R_i^{in} = a_J\left(0.5- \dfrac{3}{2N}\left((i-\dfrac{2N}{3})+\dfrac{1}{2}\right)\right).
\end{gather}
Which can be rewritten : 
\begin{gather}
|\Delta J_{2k}| \sim \dfrac{4\pi(2k+3)}{12}\dfrac{\bar{A}<P_{2k}>}{M}
\dfrac{27 \Delta r^3}{8} \nonumber \\
\times \left[\dfrac{1}{8}\sum_{i=0}^{\frac{2N}{3}}
\left( \dfrac{R_i^{ext}}{a_J}\right)^{2k+2} +
\sum_{i=0}^{\frac{N}{3}}
\left( \dfrac{R_{i-\frac{2N}{3}}^{int}}{a_J}\right)^{2k+2}\right].
\end{gather}
For $k=1$, using Eq.(\ref{simp_deltar}) and Eq.(\ref{P_fin}), we obtained : 

\begin{gather}
|\Delta J_{2}| \sim \dfrac{5\times 27\times \pi<P_{2}>}{12\times 8\times N^3}\left[\dfrac{1}{8}\left(\dfrac{4}{3}\dfrac{2\times N}{3}\right)
\dfrac{1}{5}
+\dfrac{1}{2^4}\left(\dfrac{1}{3}\dfrac{N}{3}\right)\dfrac{1}{5}\right] \nonumber \\
|\Delta J_{2}| \sim \dfrac{9}{256} \dfrac{\pi <P_2>}{N^2}.
\label{J2_hub_calc}
\end{gather}
Compared to Eq.(\ref{J_2_fin}), this is approximately twice better in terms
of errors.

\subsection{Cubic spacing}
\label{maths_cubic}

In this appendix, we calculated the error for a cubic repartition of the
spheroids,
that is $\Delta r$ depending on $i$ as $r_i = a_J - a_J \times i^3/N^3$.
%. By which we mean that the spacing between two spheroids
%will evolve as the power of three of their distance from the outside. \\

Using Eq.(\ref{J_2k_deltar}), we got:
\begin{gather}
\Delta r_i = ((i+1)^3 -i^3)  \dfrac{a_J}{N^3} = \dfrac{\Delta (i^3)}{N^2} \Delta r, 
\label{delta_ri} \\
R_i = a_J \left(1- \dfrac{i^3+(i+1)^3}{2N^3} \right).
\end{gather}
Straightforwardly, we obtained the new Eq.(\ref{J_2k_final}) :

\begin{align}
|\Delta J_{2k}^c| \simeq (2k+3)\dfrac{\pi}{12}\dfrac{<|P_{2k}|>}{N^5} \nonumber \\
\times \sum_{i=0}^{N-1} \Delta (i^3) \left(1- \dfrac{i^3+(i+1)^3}{2N^3} \right)^{2k+2},
\label{J_2k_c}
\end{align}
which gave for $J_2$ : 
\begin{equation}
|\Delta J_{2}^c| \simeq \dfrac{5 \pi <P_2>}{12N^5}
\sum_{i=0}^{N-1} \Delta (i^3) \left(1- \dfrac{i^3+(i+1)^3}{2N^3} \right)^{4}.
\end{equation}
We developed again the various terms:
\begin{gather}
\Delta (i^3) = 3i^2 + 3i + 1 \sim 3i^2,\\
P^c = \left(1- \dfrac{i^3+(i+1)^3}{2N^3} \right)^{4}, \nonumber \\
P^c = \left(1- \dfrac{1}{N^3}i^3 - \dfrac{3}{2N^3}i^2 - \dfrac{3}{2N^3}i - \dfrac{1}{2N^3} \right)^4
 \simeq \left(1-\dfrac{i^3}{N^3} \right)^4, \nonumber \\
 P^c \simeq 1 -4\dfrac{i^3}{N^3} + 6 \dfrac{i^6}{N^6} - 4 \dfrac{i^9}{N^9} + \dfrac{i^{12}}{N^{12}}, \\
 \Delta (i^3) P^c \simeq 3 \left(i^2 -4\dfrac{i^5}{N^3} + 6\dfrac{i^8}{N^6} - 4\dfrac{i^{11}}{N^9} + \dfrac{i^{14}}{N^{12}} \right).
 \label{P_cubic}
 \end{gather}
In order to calculate the sum, we needed the leading term of $\sum_{i=0}^{N-1} i^k$. With: 
\begin{gather}
(a+1)^k - a^k = \sum_{i=0}^{k-1} \dbinom{k}{i} a^i, \\
\sum_{a=1}^{N-1} \left((a+1)^k - a^k \right) = \sum_{i=0}^{k-1} 
\dbinom{k}{i} \sum_{a=1}^{N-1}a^i, \\
\text{  but the terms in the sum cancelled two by two :  } \nonumber \\
\sum_{a=1}^{N-1} \left((a+1)^k - a^k \right) = N^k - 1,  \\
\text{ so: }\,\,\,\,N^k-1 = \sum_{i=0}^{k-1} \dbinom{k}{i} \sum_{a=1}^{N-1}a^i.
\end{gather}
With this formula, it is easy to show by recurrence that $\sum_{a=1}^{N-1}a^i = O(N^{i+1})$. 
Therefore, in the equation above only the $k-1$ term 
can lead to $N^k$. Then the result: 
\begin{equation}
\sum_{a=1}^{N-1}a^{k-1} \sim \dfrac{N^k}{\dbinom{k}{k-1}} = \dfrac{N^k}{k}.
\end{equation}
From here, we obtained the leading order terms of the sum of Eq.(\ref{P_cubic}) :
\begin{gather}
\sum_{i=0}^{N-1}\Delta (i^3) P^c \sim 3 \left(\dfrac{N^3}{3} - \dfrac{4N^3}{6} + \dfrac{6N^3}{9}
- \dfrac{4N^3}{12} + \dfrac{N^3}{15} \right) = \dfrac{N^3}{5}
\end{gather}
The result is thus the same as in the linear case : 
\begin{equation}
|\Delta J_{2}^c| \sim \dfrac{\pi}{12} \dfrac{<P_2>}{N^2} .
\end{equation}

\subsection{Exponential spacing}
\label{maths_exp}
For an exponential repartition of the spheroids, we imposed :
\begin{equation}
\lambda_{i+1} = \lambda_{i} - \beta e^{i\alpha}.
\end{equation}
Then, considering the difference $\lambda_{i+1} - \lambda_{i}$
as a geometric sequence, we obtained : 
\begin{equation}
\lambda_{i} = 1 - \beta \dfrac{e^{i\alpha}-1}{e^{\alpha}-1}.
\end{equation}
Now, with the condition that the first layer (normalized to the radius of Jupiter) is 1 and the last one is $0$ : 
\begin{equation}
\lambda_{i} = 1 - \dfrac{e^{i\alpha}-1}{e^{N\alpha}-1}.
\label{lambda_exp}
\end{equation}
To determine the error we needed : 
\begin{gather}
R_{i} = 1 + \dfrac{1}{e^{N\alpha}-1} - \dfrac{e^{i\alpha}}{2(e^{N\alpha}-1)} 
-\dfrac{e^{(i+1)\alpha}}{2(e^{N\alpha}-1)}  = \nonumber \\
1 + \dfrac{1}{e^{N\alpha}-1} 
- e^{i\alpha}\dfrac{1+e^{\alpha}}{2(e^{N\alpha}-1)},  \\
\Delta r = \dfrac{e^{(i+1)\alpha}-e^{i\alpha}}{e^{N\alpha}-1} = 
e^{i\alpha}\dfrac{e^{\alpha}-1}{e^{N\alpha}-1} = \beta e^{i\alpha}.
\end{gather}
As we wanted the increment to be reasonable, $e^{N\alpha}-1$ could not be too large.
In this paper, we chose 
\begin{equation}
\alpha = \dfrac{\gamma}{N},
\end{equation}
where we varied $\gamma$ between 5 and 10, depending on how sharp we 
wanted the exponential function to be (see Figure \ref{fig:repartition}). \\

\noindent Then, using Eq.(\ref{J_2k_deltar}), we had to calculate, for $J_2$ : 
\begin{gather}
\sum_{i=0}^{N-1} (\Delta r_i)^3 R_i^4 =  \nonumber \\
\beta^3 \sum_{i=0}^{N-1} e^{3i\alpha} 
\left( \left[1 + \dfrac{1}{e^{N\alpha}-1} \right]
- e^{i\alpha}\dfrac{1+e^{\alpha}}{2(e^{N\alpha}-1)} \right)^4, \nonumber \\
\sum_{i=0}^{N-1} (\Delta r_i)^3 R_i^4 = 
\beta^3 \sum_{i=0}^{N-1} e^{3i\alpha}(\delta  - \epsilon e^{i\alpha})^4, \\
\beta = \dfrac{e^{\alpha}-1}{e^{N\alpha}-1} \text{ , }
\delta = 1 + \dfrac{1}{e^{N\alpha}-1} \text{ , }
\epsilon = \dfrac{1+e^{\alpha}}{2(e^{N\alpha}-1)}.
\end{gather}
Developing the power of 4 brackets and using the well known result for 
the sum of the terms of a geometric sequence: 

\begin{gather}
\sum_{i=0}^{N-1} (\Delta r_i)^3 R_i^4 = \beta^3 \delta^4 \left(\dfrac{e^{3N\alpha} -1}{e^{3\alpha}-1}  
- 4 \dfrac{\epsilon}{\delta} \dfrac{e^{4N\alpha} -1}{e^{4\alpha}-1}  \nonumber \right.\\\left.
+ \,6 \left(\dfrac{\epsilon}{\delta}\right)^2\dfrac{e^{5N\alpha} -1}
{e^{5\alpha}-1} -  4 \left(\dfrac{\epsilon}{\delta}\right)^3\dfrac{e^{6N\alpha} -1}
{e^{6\alpha}-1} + \right. \nonumber \\ \left. \left(\dfrac{\epsilon}{\delta}\right)^4\dfrac{e^{7N\alpha} -1} 
{e^{7\alpha}-1} \right).
\end{gather}
Here, we considered that $7\alpha \ll 1$, $e^{N \alpha} \gg 1$, which 
is a reasonable approximation in the range $\gamma = 5-10$ and $N > 512$. Then : 

\begin{gather}
\beta \sim \dfrac{\alpha}{e^{N\alpha}} \text{ , }
\delta \sim 1 \text{ , }
\epsilon \sim \dfrac{1}{e^{N\alpha}} \text{ , }
\dfrac{e^{3N\alpha} -1}{e^{3\alpha}-1} \sim \dfrac{e^{3N\alpha}}{3\alpha}, \\
\sum_{i=0}^{N-1} (\Delta r_i)^3 R_i^4 \sim 
\left(\dfrac{\alpha}{e^{N\alpha}} \right)^3 \left(\dfrac{e^{3N\alpha}}{3\alpha} -
4 \dfrac{e^{3N\alpha}}{4\alpha} +  \right. \nonumber \\ \left.
6\dfrac{e^{3N\alpha}}{5\alpha} - 4
\dfrac{e^{3N\alpha}}{6\alpha} + \dfrac{e^{3N\alpha}}{7\alpha} \right), \nonumber \\
\sum_{i=0}^{N-1} (\Delta r_i)^3 R_i^4 \sim -\dfrac{\alpha^2}{105} = -\dfrac{\gamma^2}{105 \, N^2}.
\end{gather}
We checked
numerically that our approximations yield an error allways smaller than $\approx 10\%$. \\

\noindent Implementing these results into Eq.(\ref{J_2k_deltar}) yields: 
\begin{equation}
|\Delta J_{2}^{exp}| \sim \dfrac{\pi <P_2>}{252}*\dfrac{\gamma^2}{ N^2}
\approx 0.000449 \left(\dfrac{\gamma}{N}\right)^2,
\end{equation}
with $\gamma \in [5-10]$.

\section{Neglecting the high atmsophere}
\label{app:phi_neg}

The neglected potential on a point ($\mu$, $r_j$) of the j$^{th}$ spheroid reads : 
\begin{gather}
\phi_{neg} = 4 \pi G \sum_{k=0}^{\infty} (r_j)^{2k} P_{2k}(\mu) 
\int_{a_{1bar}}^{a_{ext}}\int_{0}^{1}
\dfrac{\rho(\vec{r'})}{r'^{2k-1}} P_{2k}(\mu')d\mu' dr'.
\end{gather}

\noindent With $\rho = \rho_{ext} = cst$ and Appendix \ref{mu_legendre} :

\begin{gather}
I = \int_{a_{1bar}}^{a_{ext}}\int_{0}^{1}
\rho(\vec{r'})\dfrac{1}{r'^{2k-1}} P_{2k}(\mu')d\mu' dr' \nonumber \\ 
\simeq \rho_{ext}\int_{a_{1bar}}^{a_{ext}}\dfrac{1}{r'^{2k-1}}dr'
\int_{0}^{1} P_{2k}(\mu') (1+e^2 \mu'^2)^{k-1}d\mu'.
\end{gather}

\noindent For any $k>0$, we integrate a Legendre polynomial of degree $2k$ with
a polynomial of order $2(k-1)$. By orthogonality of the Legendre base, 
only the $k=0$ term is non zero.
Thus 
\begin{gather}
\phi_{neg} \simeq 4 \pi G \rho_{ext} \left[\dfrac{1}{2} (a_{ext}^2 - a_{1bar}^2) \dfrac{\arctan(e)}{e} \right].
\end{gather}

\noindent With $a_{ext} = a_{1bar}+\Delta a$ : 
\begin{gather}
(a_{ext}^2 - a_{1bar}^2) \simeq 2 a_{1bar}\Delta a \text{, and so}\\
\phi_{neg} \simeq 4 \pi G \rho_{ext} a_{1bar} \Delta a \dfrac{\arctan(e)}{e}.
% \dfrac{1}{a_{1bar}^{2k-2}} - \dfrac{1}{a_{ext}^{2k-2}} \simeq (2k-2)\dfrac{\Delta a}{a_{1bar}^{2k-1}}
\end{gather}
Using the exterior $e \simeq 0.38$ and defining $\bar{\rho_S}$ as the density of Jupiter if it was
a sphere of radius $a_{1bar}$, the neglected potential simply reads:
\begin{gather}
\phi_{neg} \simeq 2.87 \dfrac{G M}{a_{1bar}} \left(\dfrac{\rho_{ext}}{\bar{\rho_S}} \right)
\left(\dfrac{\Delta a}{a_{1bar}}  \right) 
\end{gather} 
This expression does not depend on $r_j$ nor on $\mu$ : the first
order error is a constant neglected potential on each spheroid.

This means that the hydrostatic condition is not perturbed by
the neglected high atmosphere, because it is only sensitive to the gradient 
of the potential. As the 1 bar pressure is set at the observed 
radius, we do not expect a change in the pressure-density profile
of the planet from the direct effect of this neglected potential.

On the other hand, this neglected potential affects the calculation of
the shapes of the spheroids. Calling $\Delta U = \phi_{neg} \times a_{1bar}/ GM$
and $U_j$ the dimensionless potential of the j$^{th}$ spheroid, we derived
the first order perturbation of equation (51) of H13 : 
\begin{gather}
-\dfrac{1}{\zeta_j (\mu)} \bigg ( \sum_{i=j}^{N-1} \sum_{k=0}^{\infty} \tilde{J}_{i,2k} 
\left(\dfrac{\lambda_i}{\lambda_j}\right)^{2k} \zeta_j(\mu)^{-2k} P_{2k} (\mu)  \nonumber \\
+\sum_{i=0}^{j-1} \sum_{k=0}^{\infty} \tilde{J'}_{i,2k} 
\left(\dfrac{\lambda_j}{\lambda_i}\right)^{2k+1} \zeta_j(\mu)^{2k+1} P_{2k} (\mu)  \nonumber \\
 + \sum_{i=0}^{j-1} \tilde{J''}_{i,0} 
\lambda_j^3 \zeta_j(\mu)^3 \bigg) + \dfrac{1}{2} \lambda_j^3 \zeta_j(\mu)^2 (1 - \mu^2) = U_{j} + \Delta U.
\label{eq:51_HM13}
\end{gather} 
For simplicity, we set $x = \bar{\zeta}_j (\mu)$ the solution without the neglected 
potential and $\Delta x$ the variation to this solution. Both $x$ and $\Delta x$ depend
on $\mu$. It was then straightforward to write Eq.\eqref{eq:51_HM13} 
at the first order : 
\begin{gather}
-\dfrac{1}{x} \bigg (- \sum_{i=j}^{N-1} \sum_{k=0}^{\infty} \tilde{J}_{i,2k} 
\left(\dfrac{\lambda_i}{\lambda_j}\right)^{2k} x^{-2k} 
\left(2k \dfrac{\Delta x}{x} \right) P_{2k} (\mu)  \nonumber \\
+\sum_{i=0}^{j-1} \sum_{k=0}^{\infty} \tilde{J'}_{i,2k} 
\left(\dfrac{\lambda_j}{\lambda_i}\right)^{2k+1} x^{2k+1} \left((2k+1) 
\dfrac{\Delta x}{x} \right) P_{2k} (\mu)  \nonumber \\+ \sum_{i=0}^{j-1} \tilde{J''}_{i,0} 
\lambda_j^3 x^3 \left(3\dfrac{\Delta x}{x} \right) \bigg) \nonumber \\
+\dfrac{1}{x} \dfrac{\Delta x}{x} \bigg ( \sum_{i=j}^{N-1} \sum_{k=0}^{\infty} \tilde{J}_{i,2k} 
\left(\dfrac{\lambda_i}{\lambda_j}\right)^{2k} x^{-2k} P_{2k} (\mu) \nonumber \\+
\sum_{i=0}^{j-1} \sum_{k=0}^{\infty} \tilde{J'}_{i,2k} 
\left(\dfrac{\lambda_j}{\lambda_i}\right)^{2k+1} x^{2k+1} P_{2k} (\mu)  \nonumber \\
 + \sum_{i=0}^{j-1} \tilde{J''}_{i,0} 
\lambda_j^3 x^3 \bigg) + 
\lambda_j^3 x^2 \left(\dfrac{\Delta x}{x} \right) (1 - \mu^2) = \Delta U.
\end{gather}
The second bracket could be simplified by the fact that $x$ satisifies equation (51)
of H13 : 
\begin{gather}
-\dfrac{\Delta x}{x} \bigg (- \sum_{i=j}^{N-1} \sum_{k=0}^{\infty} 2k \tilde{J}_{i,2k} 
\left(\dfrac{\lambda_i}{\lambda_j}\right)^{2k} x^{-2k-1} P_{2k} (\mu)  \nonumber \\
+\sum_{i=0}^{j-1} \sum_{k=0}^{\infty}(2k+1) \tilde{J'}_{i,2k} 
\left(\dfrac{\lambda_j}{\lambda_i}\right)^{2k+1} x^{2k}
P_{2k} (\mu)  \nonumber \\
+ \sum_{i=0}^{j-1} 3\tilde{J''}_{i,0} 
\lambda_j^3 x^2 \bigg)
-\dfrac{\Delta x}{x} \bigg (U_j -  \dfrac{1}{2} \lambda_j^3 x^2 (1 - \mu^2)\bigg)  \nonumber \\+ 
\left(\dfrac{\Delta x}{x} \right)\lambda_j^3 x^2  (1 - \mu^2) = \Delta U.
\end{gather}
Now, we restricted ourselves to the outermost spheroid $j=0$, $\lambda_j = 1$. 
Only the first sum  remained, and we recoginized Eq.\eqref{J} : 
\begin{gather}
\dfrac{\Delta x}{x} \bigg (\sum_{k=0}^{\infty} 2k J_{2k}^{ext} x^{-2k-1} 
P_{2k} (\mu) 
- U_j + 
\dfrac{3}{2}x^2(1 - \mu^2) \bigg)= \Delta U.
\end{gather} 
With Juno data, we know the first few $J_{2k}^{ext}$ and the contribution
of each $J_{2k}$ to the sum is strongly decreasing with $k$ so we were able to use 
only the first 4 even gravitational moments. From here, we
approximated $x$ given by the CMS program as a polynomial of order 15 (which
gave errors of a few $10^{-14}$), but allowed us to obtain as many evaluations
of $\Delta x$ as we needed. 

We noticed that around $\mu \sim 0.5$ the bracketed term tends to 0 so $\Delta x$
should not be defined. That means that at this point some neglected effect should
be taken into account. However, with the polynomial approximation for $x$, we were able to
use Riemann sum to evaluate the change on equation (40) of H13 : 
\begin{gather}
\tilde{J}_{i,2k} = -\dfrac{3}{2k+3} \dfrac{\delta \rho_i \lambda_i^3 
\int_0^1 P_{2k}(\mu) \zeta_i(\mu)^{2k+3}d\mu}{\sum_{j=0}^{N-1} \delta \rho_j
\lambda_j^3 \int_0^1 \zeta_j(\mu)^3 d\mu}.
\label{eq:delta_atmo}
\end{gather}
We found that the undefined region for $\Delta x$ has a negligible influence
on the integral because it is almost ponctual. We also found
that the change in the denominator which is, as expected, of the order
of the neglected mass, is ten times smaller than 
the change in the numerator. Therefore, the change in the numerator
dominates the uncertainties on $J_2$.
By comparing for different number of points in the Riemann sum, we obtained :
\begin{equation}
\dfrac{\Delta J_{2,0}}{J_{2,0}} \simeq 100 \times \Delta U \simeq 287 \times
\left(\dfrac{\rho_{ext}}{\bar{\rho_S}} \right)
\left(\dfrac{\Delta a}{a_{1bar}}  \right).
\end{equation}
This result has a variation of about a factor 2 (due to the bad handling
of the divergence zone) and this evaluation is rather a lower
bound for $\Delta J_{2,0}$.

Because, in the linear limit, the $2k+3$ exponent cancel the $2k+3$
in the denominator, the absolute uncertainty is approximatively
the same for all $J_k$ (decorrelated by the $P_{2k}$) so the relative uncertainty is a rapidly 
growing function of k. Typically:

\begin{gather}
\dfrac{\Delta J_{4,0}}{J_{4,0}} \simeq 1 \times 10^3 \times \Delta U \\
\dfrac{\Delta J_{6,0}}{J_{6,0}} \simeq 6 \times 10^3\times \Delta U \\
\dfrac{\Delta J_{8,0}}{J_{8,0}} \simeq 1 \times 10^5 \times \Delta U 
\end{gather}

% \begin{thebibliography}{MaBiblio}
% \bibitem[Zharkov and Trubitsyn(1978)]{Zharkov} {Zharkov}, V.~N. and {Trubitsyn}, V.~P. 1978 Physics of 
% Planetary Interiors (Tucson,AZ: Pachart)
% \bibitem[Hubbard(2012)]{Hub2012} Hubbard W. B. 2012, \apj Letters, 756, L15
% \bibitem[Hubbard(2013)]{Hub2013} Hubbard W. B. 2013, \apj , 768, 43
% \bibitem[Hubbard and Militzer(2016)]{HM2016} Hubbard W. B. \& Militzer B.  2016, \apj , 820, 80
% \bibitem[Hubbard(1975)]{Hub1975} Hubbard W. B. 1975, Soviet Astronomy , 18, 621
% \bibitem[Chabrier et al.(1992)]{Chab92} {Chabrier}, G. and {Saumon}, D. and {Hubbard}, W.~B. and Lunine, J.~I.
% 1992, \apj, 391, 817
% \bibitem[Cohen and Taylor(1987)]{Cohen87} Cohen E. R. \& Taylor B. N. 1987, Rev. Mod. Phys. 59, 1121
% \bibitem[Campbell and Synott(1985)]{Campbell85} Campbell J. K. \& Synott S. P. 1985, \aj, 364
% \bibitem[Archinal et al.(2011)]{Archinal2011}{Archinal}, B.~A. and {A'Hearn}, M.~F. and {Bowell}, E. and 
%   {Conrad}, A. and {Consolmagno}, G.~J. and {Courtin}, R. and 
%   {Fukushima}, T. and {Hestroffer}, D. and {Hilton}, J.~L. and 
%   {Krasinsky}, G.~A. and {Neumann}, G. and {Oberst}, J. and {Seidelmann}, P.~K. and 
%   {Stooke}, P. and {Tholen}, D.~J. and {Thomas}, P.~C. and {Williams}, I.~P. 2011, Celestial Mechanics
%     And Dynamical Astronomy, 109, 101
% \bibitem[Riddle and Warwick(1976)]{Riddle76} Riddle A. C. \$ Warwick J. W. 1976, Icarus, 27, 457

% \end{thebibliography}

\end{document}